\documentclass[%
 reprint,
 amsmath,amssymb,
 aps,
prb
]{revtex4-2}

\usepackage{graphicx}
\usepackage{dcolumn}
\usepackage{bm}
\usepackage{xcolor}

\usepackage{subfig}
\usepackage{float}
\usepackage{array,multirow}
\usepackage[nolist]{acronym}

\begin{acronym}
	\acro{RPA}{Random Phase Approximation}
	\acro{ERPA}{Extended Random Phase Approximation}
	\acro{QBA}{Quasi-Boson Approximation}
	\acro{RPAX}{Random Phase Approximation with Exchange}
	\acro{dRPA}{direct Random Phase Approximation}
	\acro{FCI}{Full Configuration Interaction}
	\acro{RDM}{Reduced Density Matrix}
	\acro{RDMs}{Reduced Density Matrices}
	\acro{HF}{Hartree--Fock}
	\acro{CAS}{complete active space}
	\acro{CI}{Configuration Interaction}
	\acro{NO}{Natural Orbital}
	\acro{DMET}{Density Matrix Embedding Theory}
	\acro{EwDMET}{Energy-weighted Density Matrix Embedding Theory}
	\acro{EoM}{Equation of Motion}
	\acro{SVD}{Singular Value Decomposition}
	\acro{DFT}{Density Functional Theory}
	\acro{TDM}{Transition Density Matrix}
	\acro{RI}{Resolution of the Identity}
	\acro{EDMET}{`Extended'-DMET}
	\acro{DMFT}{Dynamical Mean Field Theory}
	\acro{TDDFT}{Time-Dependent Density Functional Theory}
	\acro{DFT}{Density Functional Theory}
	\acro{PPP}{Pariser--Parr--Pople}
	\acro{MRSDCI}{multireference singles-doubles configuration-interaction}
	\acro{CO}{charge ordered}
	\acro{AFM}{anti-ferromagnetic}
	\acro{EHM}{extended Hubbard model}
	\acro{BOW}{bond order wave}
	\acro{DMRG}{density matrix renormalisation group}
	\acro{DDMRG}{dynamical DMRG}
	\acro{SDW}{spin density wave}
	\acro{CDW}{charge density wave}
	\acro{EDMFT}{extended DMFT}
	\acro{DCA}{Dynamical Cluster Approximation}
	\acro{DIIS}{direct inversion in the iterative subspace}
\end{acronym}

\usepackage{algorithm}
\usepackage{algpseudocode}

\newcommand{\toadd}[1]{{#1}}
\newcommand{\vthree}[1]{{#1}}

\begin{document}


\title{Extending Density Matrix Embedding: A Static Two-Particle Theory}

\author{Charles J. C. Scott}
 \email{cjcargillscott@gmail.com}
\author{George H. Booth}%
 \email{george.booth@kcl.ac.uk}
\affiliation{%
 Department of Physics, King’s College London, The Strand, London WC2R 2LS, United Kingdom
}%

\date{\today}

\begin{abstract}
We introduce Extended Density Matrix Embedding Theory (EDMET), a static quantum embedding theory explicitly self-consistent with respect to \toadd{local} two-body \toadd{physics}. This overcomes the biggest practical and conceptual limitation of more traditional one-body embedding methods, namely the lack of screening and treatment of longer-range \toadd{interactions}.
This algebraic zero-temperature embedding augments a \toadd{local interacting cluster model} with a minimal number of bosons from \toadd{a description of the full system correlations via the} random phase approximation, and admits an analytic approach to build a self-consistent Coulomb-exchange-correlation kernel. For extended Hubbard models with non-local interactions, this leads to the accurate description of phase transitions, static quantities and dynamics. We also move towards {\em ab initio} systems via the Parriser--Parr--Pople model of conjugated coronene derivatives, finding good agreement with experimental optical gaps.
\end{abstract}

\maketitle

\newcounter{dummy} 
\newcommand{\diff}{\mathop{}\!\mathrm{d}} 
\newcommand{\Supp}{\mathop{}\!\mathrm{Supp}}
\newcommand{\deriv}[3][]{
\frac{\diff^{#1}#2}{\diff #3^{#1}}}
\newcommand{\pderiv}[3][]{%
\frac{\partial^{#1}#2}
{\partial #3^{#1}}}
\newcommand{\vect}[1]{\mathbf{#1}} 
\newcommand{\mat}[1]{\mathbf{#1}}
\newcommand{\Pcl}{\mat{P}_{\mathrm{cl}}}
\newcommand{\Pferm}{\mat{P}_{\mathrm{cl,f}}}
\newcommand{\PRPA}{\mat{P}_{\mathrm{RPA}}}
\newcommand{\Pfrag}{\mat{P}_{\mathrm{frag}}}
\newcommand{\blankpage}{\newpage\hbox{}\thispagestyle{empty}\newpage}
\newcommand{\hairsp}{\hspace{1pt}}
\newcommand{\ie}{\textit{i.\hairsp{}e.}\xspace}
\newcommand{\Ie}{\textit{I.\hairsp{}e.}\xspace}
\newcommand{\eg}{\textit{e.\hairsp{}g.}\xspace}
\newcommand{\Eg}{\textit{E.\hairsp{}g.}\xspace}
\newcommand{\vs}{\textit{vs.}\xspace}
\newcommand{\abs}[1]{\left|#1\right|}
\newcommand{\bra}[1]{\left\langle#1\right|}
\newcommand{\ket}[1]{\left|#1\right\rangle}
\newcommand{\braket}[2]{\left\langle#1|#2\right\rangle}
\newcommand{\order}[1]{\mathcal{O}\left(#1\right)}
\newcommand{\normord}[1]{n[#1]}
\newcommand{\normordf}[1]{\{#1\}}
\newcommand{\normordop}[1]{\hat{#1}_{\textrm{N}}}
\newcommand{\commu}[2]{\left[#1,#2\right]}

\newcommand{\crea}[1]{a_{#1}^\dagger}
\newcommand{\annih}[1]{a_{#1}}
\newcommand{\bcrea}[1]{b_{#1}^\dagger}
\newcommand{\bannih}[1]{b_{#1}}
\newcommand{\rpamom}[1]{\mat{\eta}^{(#1)}}
\newcommand{\Xbos}{\mat{X}_{\mathrm{bos}}}
\newcommand{\Ybos}{\mat{Y}_{\mathrm{bos}}}
\newcommand{\ghb}[1]{{\color{red}GHB: #1}}
\newcommand{\cjcs}[1]{{\color{cyan}CJCS: #1}}

\begin{acronym}
	\acro{RPA}{Random Phase Approximation}
	\acro{ERPA}{Extended Random Phase Approximation}
	\acro{QBA}{Quasi-Boson Approximation}
	\acro{RPAX}{Random Phase Approximation with Exchange}
	\acro{dRPA}{direct Random Phase Approximation}
	\acro{FCI}{Full Configuration Interaction}
	\acro{RDM}{Reduced Density Matrix}
	\acro{RDMs}{Reduced Density Matrices}
	\acro{HF}{Hartree--Fock}
	\acro{CAS}{complete active space}
	\acro{CI}{Configuration Interaction}
	\acro{NO}{Natural Orbital}
	\acro{DMET}{Density Matrix Embedding Theory}
	\acro{EwDMET}{Energy-weighted Density Matrix Embedding Theory}
	\acro{EoM}{Equation of Motion}
	\acro{SVD}{Singular Value Decomposition}
	\acro{DFT}{Density Functional Theory}
	\acro{TDM}{Transition Density Matrix}
	\acro{RI}{Resolution of the Identity}
	\acro{EDMET}{`Extended'-DMET}
	\acro{DMFT}{Dynamical Mean Field Theory}
	\acro{TDDFT}{Time-Dependent Density Functional Theory}
	\acro{DFT}{Density Functional Theory}
	\acro{PPP}{Pariser--Parr--Pople}
	\acro{MRSDCI}{multireference singles-doubles configuration-interaction}
	\acro{CO}{charge ordered}
	\acro{AFM}{anti-ferromagnetic}
	\acro{EHM}{extended Hubbard model}
	\acro{BOW}{bond order wave}
	\acro{DMRG}{density matrix renormalisation group}
	\acro{DDMRG}{dynamical DMRG}
	\acro{SDW}{spin density wave}
	\acro{CDW}{charge density wave}
	\acro{EDMFT}{extended DMFT}
	\acro{DCA}{Dynamical Cluster Approximation}
	\acro{DIIS}{direct inversion in the iterative subspace}
\end{acronym}


\toadd{\section{Introduction}}

Strongly correlated materials exhibit some of the most fascinating and technologically important emergent phenomena in condensed matter \cite{Kent2018,Wagner2014}. However, due to the presence of a non-perturbative quantum many-body problem at their heart, are notoriously difficult to simulate accurately\citep{Alling2010,Basov2011}. Furthermore, strong correlations in these materials can rarely be considered in isolation, and so a faithful description of non-local and high-energy interactions is essential. These couple to the strongly correlated bands which introduces a screening of the low-energy interactions to qualitatively change the resulting physics \cite{Boehnke2016}.
Quantum cluster methods, such as dynamical mean-field theory (DMFT), have emerged in the last couple of decades as the dominant approach to extend simulation capabilities to strongly correlated materials\citep{Sun2016}. These operate via a self-consistent mapping of the system to an auxiliary embedded quantum cluster problem, defined by a fragment of the system coupled to a `bath', designed to represent the renormalized propagation of local one-body quasiparticles through the system\cite{Georges1996}.
This cluster problem can then be solved to high accuracy with various techniques \cite{Hirsch1986,Werner2006,Werner2006a,Zgid2012,Lu2014,Go2017,Zhu2019}, and the resultant local properties used to self-consistently update the overall system dynamical Green's function. 

This DMFT approach has proved a highly effective approximation when a system is dominated by interactions which are local to the choice of fragment \toadd{(often denoted the `impurity')} \citep{Savrasov2001}.
However, DMFT neglects \toadd{dynamic} screening effects and other collective phenomena which derive from longer-range non-local interactions, and which cannot be ignored within most realistic materials \cite{Boehnke2016}. To account for this, a variety of uncontrolled approximations are often used to statically vary the effective interactions of the cluster model to account for these non-local interactions on the local physics\citep{Gunnarsson1989,Gunnarsson1990, Kotani2000,Aryasetiawan2004},
or assume an independent additive contribution of the non-local interactions to the strongly correlated \toadd{local physics of the} fragment \cite{Zhu2021}.
A rigorous approach to couple these effects necessitates the introduction of a frequency-dependent `screened' Coulomb interaction in the cluster model, \toadd{which is denoted} `Extended'-DMFT\citep{Almbladh1999,Biermann2002,Haule2007,Ayral2013,Choi2016,Lechermann2017,Medvedeva2017,Tomczak2017,VanLoon2018,Rohringer2018}. \toadd{This approach involves explicit mapping of two-body quantities between the fragment and full system}, and is more involved than the comparatively simple one-body embedding of DMFT. Despite this, it has demonstrated that it can qualitatively describe strongly correlated materials in a fully ab initio approach, \toadd{fully coupling both one-body and two-body physics.} \citep{Petocchi2020}.
However, the requirement to find and self-consistently solve the auxiliary problem with a fully frequency-dependent Coulomb interaction has limited the scope of the technique and accessible fragment sizes.

\ac{DMET}\citep{Knizia2012,Knizia2013} takes a different approach to quantum embedding, neglecting a large portion of the dynamical information in the hybridization, bath construction and self-consistency, to work in an entirely static framework with a compact and algebraic bath construction. These simplifications admit an explicit hamiltonian formulation and efficient, zero-temperature wave function methods for solving the cluster \cite{Chen2014,Holmes2016a,Zheng2017b,Pham2018,Cui2020}, with its compact size and static nature enabling large fragments to be considered and avoid analytic continuation.\citep{Cui2020}
The recently introduced {\em energy-weighted} density matrix embedding allows for this neglected dynamical character of the one-body properties to be systematically restored to the theory at the cost of an enlarged bath space, directly connecting to the method back to a static analog of DMFT \cite{Fertitta2018,Fertitta2019,Sriluckshmy2021}.
However, these approaches again neglect non-local screening effects and a controllable and systematic inclusion of two-body interactions between the fragment and wider material. DMET on real materials has relied on the fragment being large enough to saturate these important non-local interactions, which can be challenging.\citep{LeBlanc2015,Zheng2017a}

In this \toadd{work} we demonstrate an alternative approach, inspired by EDMFT. We rigorously extend DMET to include the coupling to non-local correlations, screening and long-range collective phenomena present in real materials, maintaining an efficient static framework and permitting an algebraic self-consistent approach for \toadd{both one- and} two-body \toadd{fragment-local} properties. In keeping with precedent, we call this approach \ac{EDMET}.

To achieve this, we formulate an interacting cluster \toadd{model} which exactly and algebraically reproduces the lowest-order local spectral moments of the two-particle density-density (dd) response of the full system at the level of the Random Phase Approximation (RPA). \toadd{This is analogous to the bath construction criteria of DMET, where a bath is found which exactly and algebraically reproduces the lowest-order local spectral moments of the one-particle Greens function (the one-body reduced density matrix) at a mean-field level.} Furthermore, a well-defined self-consistency is formulated to ensure these correlated local two-body properties are included back in the full system RPA description. In this initial work we focus on self-consistency in the zeroth and first moments of this dd-response, \toadd{ensuring convergence of} large parts of the local two-body reduced density matrix and beyond, and show that this can allow for accurate two-particle quantities and phase diagrams of both lattice models with non-local Coulomb terms, as well the optical gaps of Pariser-Parr-Pople (PPP) models of aromatic hydrocarbons.

\toadd{
\section{EDMET Theory}
\subsection{Defining the RPA}}

The \ac{RPA} resums bubble diagrams to infinite order, constructing chains of virtual electron polarization events coupled by the Coulomb interaction for all time-orderings \cite{Gell-Mann1957}. This defines the dd-response of the system, $\chi(\omega)$, mediated by an effective screened Coulomb interaction which couples all many-body and long-range density fluctuations, including correlated collective excitations such as plasmons \citep{Eguiluz1983,Li1992,GarciaDeAbajo1993,Polini2008,Ichikawa2016}. This physics dominates in \toadd{correlated} systems with a large polarizability or gapless systems\citep{Chen2017}, and is a standard choice for screening interactions of real materials, \toadd{such as in the $GW$ or constrained-RPA methods \citep{Ren2012,Hedin1965,Springer1998,Aryasetiawan2004,Miyake2008,Miyake2009,Jiang2010, Sasioglu2011}.
    The dd-response of the RPA can be defined by an irreducible polarizability, ${\bf P}$, and interaction kernel, ${\mathcal{K}}$, which couples these irreducible particle-hole excitations. In the `direct' RPA, this interaction kernel is just the Coulomb interaction (neglecting exchange), ${\bf v}$, resulting in a construction of the RPA dd-response as
    \begin{equation}
        \begin{aligned}
            \chi_{\text{RPA}}(\omega) = ({\bf P}(\omega)^{-1} - {\bf v})^{-1} . \label{eq:ddresponse}
        \end{aligned}
    \end{equation}
    In the EDMET procedure we will describe, quantities defining the polarizability and interaction kernel for the full system RPA are self-consistently updated, in order to match the correlated dd-response moments from the local cluster model (including the local exchange).
    
    \toadd{For the bath construction, we turn to the Casida formulation of RPA, which can be derived from Eq.~\ref{eq:ddresponse} \citep{Casida1995,Casida1998,Furche2001a}}. The Casida equation formulates the RPA as a generalized eigenvalue problem, as
    \begin{equation}
        ({\Lambda} - \Omega_n \mat{\Delta} )\ket{\vec{X}_n,\vec{Y}_n} = 0
		\label{eq:Casida-eq}
		\end{equation} 
		where
		\begin{equation}
		\begin{aligned}
		\mat{\Delta} &= \left(
		\begin{aligned}
		& \mat{1} & \mat{0} & \\
		& \mat{0} & \mat{-1} & 
		\end{aligned}
		\right)\\
		\mat{\Lambda} &=\left(
		\begin{aligned}
		&\mat{A} &\mat{B}& \\
		&\mat{B} &\mat{A}&
		\end{aligned}
		\right)\\
		A_{ia,jb} &= \left(\epsilon_a - \epsilon_i\right) \delta_{ij}\delta_{ab} + \braket{ib}{aj} = \varepsilon + \mathcal{K}_A \\
		B_{ia,jb} &= \braket{ij}{ab} = \mathcal{K}_B.
		\end{aligned}
		\label{eq:Casida_definitions}
	\end{equation}
    In these equations, $i,j\ldots$ ($a,b\ldots$) label hole (particle) spin-orbitals respectively, $\epsilon_i$ is the energy of orbital $i$, and $\langle ij|ab\rangle$ denote the standard Coulomb integrals, which are used when the interaction kernel is simply the Coulomb interaction. These interaction kernels can however be more general, to include effects beyond the RPA (as will be exploited later for the self-consistent embedding), and therefore we also denote them as general two-body kernels, $\mathcal{K}$, which can be different for the $\mat{A}$ and $\mat{B}$ blocks, where $\mathcal{K}_A$ represents the interaction coupling excitations together, and $\mathcal{K}_B$ represents the interaction coupling an excitation and a de-excitation (which are the same when the interaction is just the Coulomb form given above).
    The eigenvalues of the Casida equation, $\Omega_n$, describe neutral excitation energies (the poles of $\chi(\omega)$) while the $\mat{X}$ and $\mat{Y}$ eigenvectors define the coefficients of these (quasi-bosonic) excitations and de-excitations respectively in the particle-hole basis. The full RPA dd-response in this basis can then be defined as
    \begin{equation}
		\mat{\chi}_\text{RPA}(\omega) =
		-
		\left(
		\begin{aligned}
		\mat{X} \hspace{10pt}& \mat{Y} \\
		\mat{Y} \hspace{10pt}& \mat{X}
		\end{aligned}
		\right)
		\left(
		\begin{aligned}
		\mat{\Omega}- \omega \mat{I}\hspace{10pt} & \mat{0} \\
		\mat{0} \hspace{10pt} & -\mat{\Omega}- \omega \mat{I}
		\end{aligned}
		\right)^{-1}
		\left(
		\begin{aligned}
		&\mat{X} & -&\mat{Y} \\
		&\mat{Y} & -&\mat{X}
		\end{aligned}
		\right)^T, \label{eq:rpaddresponse}
		\end{equation}
	where $\mat{\Omega}$ is the diagonal matrix of positive-frequency RPA excitations.}
	
	\toadd{In moving towards a static self-consistent embedding, we focus instead on the moments of the dd-response spectrum, where we define the
    $m^\textrm{th}$-order dd-response moment as
    \begin{equation}
    \begin{aligned}
    \chi^{(m)}_{pqrs}=-\frac{1}{\pi} \textrm{Im} \int_0^{\infty} \omega^m \mat{\chi}_{pqrs}(\omega) d\omega,
    \end{aligned}
    \end{equation}
    for $m \geq 0$, and where $p,q,\ldots$ label general orbitals in an arbitrary basis.
    These static quantities can in general also be defined in terms of expectation values of a wave function, as
    \begin{equation}
        \begin{aligned}
    \chi^{(m)}_{pqrs}=\langle {\hat c_q}^{\dagger} {\hat c_p} ({\hat H}-E_0)^m {\hat c_r}^{\dagger} {\hat c}_s \rangle -\langle {\hat c}_q^{\dagger} {\hat c}_p \rangle \langle {\hat c}_r^{\dagger} {\hat c}_s \rangle \delta_{m,0}, \label{eq:ExpectMoms}
    \end{aligned}
    \end{equation}
    where it can be seen that the $m=0$ moment contains the two-body reduced density matrix of the system.
	We will also find it useful to define an additional quantity, 
    \begin{equation}
        \begin{aligned}
            \eta^{(m)}_{iajb} = \chi^{(m)}_{iajb} + \chi^{(m)}_{aijb} + \chi^{(m)}_{iabj} + \chi^{(m)}_{aibj} ,
        \end{aligned}
    \end{equation}
    where both particle-hole excitation and de-excitation contributions to the dd-response moments are combined. At the level of \ac{RPA}, $\eta^{(m)}$ can be found as
    \begin{equation}
        \mat{\eta}^{(m)}_{\text{RPA}} = \left(\mat{X}+\mat{Y}\right) \mat{\Omega}^m \left(\mat{X}+\mat{Y}\right)^T.
    \end{equation}
    For the RPA, this quantity fully characterizes the two-point charge dd-response moments in any basis, $\chi^{(m)}_{ppqq}$. Defining this basis by the transformation matrices $C_{ip}$ and $C_{ap}$ for the hole and particle states respectively, $\chi^{(m)}_{ppqq}$ can be found from $\eta^{(m)}$ in the RPA approximation as
    \begin{equation}
        \chi^{(m)}_{ppqq} = C_{ip}C_{ap}[\mat{\eta}^{(m)}_{\text{RPA}}]_{ia,jb}C_{jq}C_{bq}. \label{eq:RPAlocalmom}
    \end{equation}
    Additionally, the form of the RPA imposes the following structure on the first two dd-moments 
    \citep{Casida1995,Furche2001a},
    \begin{equation}
        \mat{A}-\mat{B} = \rpamom{1} = \rpamom{0} \left(\mat{A}+\mat{B}\right) \rpamom{0}, \label{eq:zeromom}
    \end{equation}
    with the higher moments also constrained by a further recursive relationship
    \begin{align}
        \rpamom{m} &= \left(\mat{A}-\mat{B}\right) \left(\mat{A}+\mat{B}\right) \rpamom{m-2} \\
        &= \left[\rpamom{0} \left(\mat{A}+\mat{B}\right)\right]^{m} \rpamom{0}.
        \label{eq:highermoms_recursion}
    \end{align}}
    This follows from the closure and orthonormality relations of the \ac{RPA} solutions, $(\mat{X}+\mat{Y})(\mat{X}-\mat{Y})^T=(\mat{X}+\mat{Y})^T(\mat{X}-\mat{Y})=\mat{I}$. \toadd{If all $\order{N^2}$ non-vanishing moments are determined, this fully defines the \ac{RPA} dd-response.}
    \toadd{$\rpamom{m}$ can be block-diagonalised into either spin-blocked or, in the closed-shell case, separate singlet and triplet contributions, as is common in spin-orbital based \ac{RPA} approaches.\citep{Angyan2011}
    Only the component between same-spin excitations can contribute to the charge dd-response, so here we will assume only the spin-blocking of this quantity.
    Formulation of an equivalent relation for the spin-density response is the subject of current research.}

\toadd{\subsection{Bosonic bath construction from RPA}}
    \label{subsec:clusconstruct}
    In constructing our cluster, we start from the interacting bath \ac{DMET} construction, where the number of (fermionic) bath orbitals is bounded by the number of fragment orbitals, $n_f$.\citep{Knizia2012,Knizia2013,Wouters2016}
    This construction matches the zeroth and first single-particle spectral moments at the mean field level \toadd{(the fragment density matrix and fock matrix projection)}, and ensures that the entangled orbitals of the cluster can be rotated into a particle/hole basis \toadd{which is exactly spanned by the corresponding full system particle/hole basis respectively. \cite{Wouters2016,Sriluckshmy2021}. We will refer to this space as the `fermionic cluster', since the final (desired) cluster space will include additional (bosonic) bath degrees of freedom.}

    \toadd{A sufficient condition for the reproduction of the full system} $\rpamom{0}$ RPA dd-response moment in the fermionic cluster, is that the components of $\mat{X}$ and $\mat{Y}$ excitation coefficients match \toadd{after an RPA calculation within this cluster space.} 
    \toadd{Due to the fact that both the cluster Fock and Coulomb terms are preserved by the interacting bath DMET cluster construction, it will also lead to an exact reproduction of $\rpamom{1}$, due to Eq.~\ref{eq:zeromom}.}
    We can \toadd{satisfy this matching condition for the components of ${\mat X}$ and $\mat{Y}$ in the fermionic cluster} exactly, algebraically and without modification of the DMET fermionic cluster space hamiltonian, by introducing an additional {\em bosonic} bath to the cluster, which couples to these fermionic (de)excitations in the RPA equations.
    \toadd{The required fermion-boson couplings and bosonic frequencies can be constructed analytically using only the RPA solution in the full space and the projection defining the fermionic cluster. These additional bosonic bath states will represent the RPA excitations and deexcitations that couple to the environment, and are required to ensure a matching of the fermionic cluster space dd-response moments at the level of RPA between the cluster and full system solutions.}
    
    
    We first define a projector onto the fermionic cluster single-particle (irreducible) (de)excitation space as $\Pferm$. We then consider a \toadd{matrix, $\PRPA$, which defines a minimal space of {\em relevant} RPA excitations. These relevant RPA excitations are constructed as a contracted set of all RPA excitations, where the minimal number of (de)excitations have a non-zero component in the fermionic excitation space. This can be found from the non-null space of full system \ac{RPA} (de)excitations once projected into this fermionic cluster.} 
    This $\PRPA$ projector therefore maps between the \ac{RPA} excitations of the full space and a desired minimal space which will define the final cluster, with the latter of maximum dimension $2n_f^2$. $\PRPA$ is found \toadd{via SVD} from the union of the image spaces of $\Pferm \mat{X}$ and $\Pferm \mat{Y}$\toadd{, which projects out the null space of RPA excitations which have no component in the fermionic cluster space}. We can then find the components of these desired excitations and de-excitations in the fermionic cluster excitation space as $\mat{X}_{\mathrm{cl,f}} = \Pferm \mat{X} \PRPA$ \toadd{and $\mat{Y}_{\mathrm{cl,f}} = \Pferm \mat{Y} \PRPA$ respectively}.
    \toadd{However, these relevant RPA excitations will also have components in the environment to this fermionic cluster ph-space, characterized by the components $\mat{X}_{\mathrm{cl,b}}$ and $\mat{Y}_{\mathrm{cl,b}}$, which need to be found. Taking both components together, these will define the additional bosonic-like quasi-excitations which make up the additional bath states of the full cluster and define the long-range excitation character, ensuring that the RPA full cluster excitations have the appropriate projection into the fermionic portion of the cluster.} 
    
    \toadd{These environmental components in the bosonic part of the final cluster can} be found from the closure and orthogonality constraints of an RPA excitation mainfold, defining the relations
    \begin{align}
        \mat{S}&=\mat{I} - (\mat{X}_{\mathrm{cl,f}}-\mat{Y}_{\mathrm{cl,f}})^T(\mat{X}_{\mathrm{cl,f}}+\mat{Y}_{\mathrm{cl,f}}) \label{eq:BosBathA} \\
        &=(\mat{X}_{\mathrm{cl,b}}-\mat{Y}_{\mathrm{cl,b}})^T(\mat{X}_{\mathrm{cl,b}}+\mat{Y}_{\mathrm{cl,b}}). \label{eq:BosBathB}
    \end{align}
    \toadd{The non-symmetric matrix $\mat{S}$ can be built from Eq.~\ref{eq:BosBathA}, and diagonalized. Its eigendecomposition can subsequently be used to construct $\mat{X}_{\mathrm{cl,b}}$ and $\mat{Y}_{\mathrm{cl,b}}$ from Eq.~\ref{eq:BosBathB}, therefore fully defining the components of the relevant RPA (de)excitations in the fermionic cluster and its environment.}
    
    While the dimension of $\mat{S}$ is $2 n_f^2$, constraints on its rank mean that the resulting number of cluster bosons is upper bounded by the number of \toadd{same-spin ph-excitations within the fermionic cluster, $\frac{1}{2}n_f^2$, for all interaction kernels considered in this work, with Appendix~\ref{App:Bounds} formally proving this bound}.
    \toadd{With the definition of these bosonic (de)excitations which need to augment the fermionic cluster and couple to the fermionic particle-hole (de)excitations, we can find the Hamiltonian that results from this coupling, via a projection of the full system RPA hamiltonian into this fermion $\oplus$ boson cluster space. This results in couplings between the bosons and fermionic ph excitations ($V$), as well as the bosonic frequencies ($\omega$) suitable for the cluster solver. These are found as}
    \begin{align}
        V_{ia,n} = V_{ai,\bar{n}} &= \phantom{-}\mat{X}_{\mathrm{cl,f}}\mat{\Omega} \mat{X}_{\mathrm{cl,b}}^T + \mat{Y}_{\mathrm{cl,f}}\mat{\Omega}\mat{Y}_{\mathrm{cl,b}}^T & & \\
        V_{ai,n} = V_{ia,\bar{n}} &= -\mat{X}_{\mathrm{cl,f}}\mat{\Omega} \mat{Y}_{\mathrm{cl,b}}^T - \mat{Y}_{\mathrm{cl,f}}\mat{\Omega}\mat{X}_{\mathrm{cl,b}}^T & & \\
        \omega_{m,n} = \omega_{\bar{m},\bar{n}} &= \phantom{-}\mat{X}_{\mathrm{cl,b}}\mat{\Omega} \mat{X}_{\mathrm{cl,b}}^T + \mat{Y}_{\mathrm{cl,b}}\mat{\Omega}\mat{Y}_{\mathrm{cl,b}}^T & & \\
        \omega_{m,\bar{n}} = \omega_{\bar{m},n} &= -\mat{X}_{\mathrm{cl,b}}\mat{\Omega} \mat{Y}_{\mathrm{cl,b}}^T - \mat{Y}_{\mathrm{cl,b}}\mat{\Omega}\mat{X}_{\mathrm{cl,b}}^T & &,
    \end{align}
    where \toadd{$n, m$ indices refer to bosonic excitations, and $\bar{n}, \bar{m}$ indices refer to bosonic de-excitations. These couplings and frequencies are then transferred over from the RPA hamiltonian in the space of ph (de)excitations, to a second quantized cluster hamiltonian where all fermionic excitations in the cluster are considered, leading to}
    \begin{equation}
    \begin{aligned}
        \hat{H}_{\text{cl}} = &\hat{H}_\text{elec} + \sum_{ia,n} V_{ia,n} \left(\hat{c}^{\dagger}_{a} \hat{c}_{i} \hat{a}_n^\dagger + h.c. \right) + V_{ai,n} \left(\hat{c}^{\dagger}_{i} \hat{c}_{a} \hat{a}_n^\dagger + h.c. \right)
        \\
        &+ \frac{1}{2}\sum_{nm} \omega_{nm} (\hat{a}^\dagger_n\hat{a}_m + h.c.) + \frac{1}{2}\sum_{nm} \omega_{\bar{n}m} (\hat{a}^\dagger_n\hat{a}^\dagger_m + h.c.), \label{eq:sqham}
    \end{aligned}
    \end{equation}
    where $\hat{H}_\text{elec}$ is the standard DMET interacting-bath electronic cluster hamiltonian. A final symplectic Bogoliubov transformation can remove all coupling terms between the bosons, and \toadd{gives a final correlated cluster Hamiltonian of the form
    \begin{equation}
        \begin{aligned}
            \hat{H}_{\text{cl}} = &\hat{H}_\text{elec} + \sum_{ia,n} {\tilde V}_{ia,n} \left(\hat{c}^{\dagger}_{a} \hat{c}_{i} \hat{ a}_n^\dagger + h.c. \right) + \\
            &+ {\tilde V}_{ai,n} \left(\hat{c}^{\dagger}_{i} \hat{c}_{a} \hat{a}_n^\dagger + h.c. \right) + \sum_n {\tilde \omega}_n \hat{a}_n^\dagger \hat{a}_n , \label{eq:secquantham}
        \end{aligned}
    \end{equation}
    where ${\tilde V}_{ia,n}$ and ${\tilde \omega_n}$ refer to quantities with respect to these rotated and decoupled bosonic degrees of freedom ($n$) resulting from the Bogoliubov transformation, with $i$ and $a$ remaining the labels for the hole and particle fermionic spaces respectively.} 
    
    \toadd{In summary, we defined the minimal space of `relevant' RPA (de)excitations, which couple the irreducible particle-hole excitations from the fermionic cluster to the environment. By treating the environmental portion of these excitations as bosons and combining them with the standard fermionic cluster space of DMET, we ensure that the cluster RPA excitations are equivalent to the projection of the full system RPA (de)excitations into the fermionic cluster. This ensures that the $\rpamom{0}$ and $\rpamom{1}$ quantities are conserved between the cluster and full system solutions. Note that we are matching the full fermionic cluster moments, rather than just the fragment projection of these quantities. The hamiltonian which couples these bosonic modes to the fermionic excitations can be found from projecting the RPA hamiltonian into the bosonic bath space.}
    This resulting coupled electron-boson cluster of Eq.~\ref{eq:secquantham} rigorously ensures that an RPA calculation in this cluster (which is now independent of the size of the full system) would result in a $\chi(\omega)$ whose $\rpamom{0}$ and $\rpamom{1}$ exactly matches the projection of these RPA moments from the full system. \toadd{Finally, we note that the criteria of matching the projection of both the full system ${\mat X}$ and ${\mat Y}$ in the cluster was sufficient, but not necessary for the reproduction of the $\rpamom{0}$ and $\rpamom{1}$ moments in the bath construction. In future work, we will demonstrate an $\mathcal{O}(N^4)$ lower-scaling alternative approach to directly target these quantities in the construction of an appropriate bosonic bath space, since obtaining the full system $\mat{X}$ and $\mat{Y}$ matrices scales as $\mathcal{O}(N^6)$ where $N$ is the number of degrees of freedom in the full system.} 

\toadd{\subsection{Self-consistency in EDMET}}

Once we have found a second-quantized local fermion-boson interacting cluster of the form in Eq.~\ref{eq:secquantham}, it can be solved with a `high-level' correlated method, which in this work is performed via exact diagonalization. To ensure this is tractable, we truncate the Hilbert space with a restriction to only three bosonic occupations in each mode, except near phase transitions in one dimension where an occupancy of four was required to ensure convergence, \toadd{as the bosons representing the longer-ranged interactions were more strongly coupled.}

From this solution, we compute the zeroth and first two-point dd-response moments over the local fragment space, $(\chi_{\textrm{HL}}^{(m)})_{pp,qq}$, \toadd{as defined in Eq.~\ref{eq:ExpectMoms} with $m=0$ and $1$}. 
    The aim is to algebraically define an updated \toadd{interaction kernel in the cluster space, $\mathcal{K}_A$ and $\mathcal{K}_B$. This new kernel can ensure that this high-level} description of the fragment-local dd-response moments will be rigorously reproduced via a subsequent RPA in the cluster, including the local exchange and correlation effects which are included from the high-level solver to all orders. \toadd{This can self-consistently improve the original RPA solution, more accurately describing the coupling between charge fluctuations in the full system, by including this local exchange-correlation from the correlated solver and appropriately matching the descriptions between the two levels of theory.} 
    
    Defining a \toadd{rotation} from the cluster ph-excitation space to the local fragment sites, $\left(\Pfrag\right)_{pp,ia} = C_{pi}C_{pa}$ where $\mat{C}$ are the coefficients of the cluster particle and hole states, we construct composite moments from $\chi_{\textrm{HL}}$ and the cluster RPA description of the local moment, as
    \begin{equation}
        \rpamom{m}_\textrm{comp} = \rpamom{m}_\textrm{RPA} + \Pfrag^+\left(\chi_\textrm{HL}^{(m)} - \Pfrag\rpamom{m}_\textrm{RPA}\Pfrag^T\right)\left(\Pfrag^+\right)^T,
    \end{equation}
    \toadd{where $\Pfrag\rpamom{m}_\textrm{RPA}\Pfrag^T$ represents the two-point local dd-response from the RPA (as given in Eq.~\ref{eq:RPAlocalmom}) which is removed from the local description and replaced by the high-level fragment-local counterpart. $\Pfrag^{+}$ represents the pseudo-inverse of the rotation matrix (since this is not square) in order to project back into the cluster ph-space from the local fragment space.}
    This \toadd{composite quantity} allows us to invert Eq.~\eqref{eq:zeromom}, to define a non-local, static coulomb-exchange-correlation kernel of the cluster RPA hamiltonian \citep{Botti2007}, giving
    \begin{align}
        \mathcal{K}^{\textrm{cl}}_A &= \frac{1}{2} \left( {\rpamom{0}_\textrm{comp}}^{-1} \rpamom{1}_\textrm{comp} {\rpamom{0}_\textrm{comp}}^{-1} + \rpamom{1}_\textrm{comp}\right) - \mat{\varepsilon}. \\
        \mathcal{K}^{\textrm{cl}}_B &= \frac{1}{2} \left( {\rpamom{0}_\textrm{comp}}^{-1} \rpamom{1}_\textrm{comp} {\rpamom{0}_\textrm{comp}}^{-1} - \rpamom{1}_\textrm{comp}\right)
    \end{align}
    The component of this induced kernel in the bosonic space of the cluster is neglected, since this would result in double counting within the self-consistency.
    The magnitude of these neglected terms reduces during self-consistency, though will not necessarily vanish at convergence.
    
    The overall long-range interaction kernels defining the next iteration lattice RPA, is then constructed via a democratic partitioning between all clusters, as modifications within the fermionic bath space can extend between fragment spaces \cite{Wouters2016,Wu2019}.
    \toadd{For this, we define $\Delta\mathcal{K}^{\textrm{cl},x}$ as the change in the overall combined interaction kernel (including excitations and deexcitations) from the original bare coulomb interaction, $\mat{v}$, in each cluster $x$.
    We can express all of the nonzero elements (in the cluster particle-hole basis) as
    \begin{equation}
    \begin{aligned}
        \Delta\mathcal{K}^{\textrm{cl},x}_{iajb} &= \Delta\mathcal{K}^{\textrm{cl},x}_{aibj} &= \left(\mathcal{K}^{\textrm{cl},x}_A\right)_{iajb} - v_{iajb}\\
        \Delta\mathcal{K}^{\textrm{cl},x}_{iabj} &= \Delta\mathcal{K}^{\textrm{cl},x}_{aijb} &= \left(\mathcal{K}^{\textrm{cl},x}_B\right)_{iajb} - v_{iajb}.
    \end{aligned}
    \end{equation}
    The equations above make it clear that local exchange effects are introduced into the interaction kernel, since $\mathcal{K}_{iajb}$ (coupling excitations) is no longer equivalent to $\mathcal{K}_{iabj}$ (coupling an excitation and de-excitation), as is the case for just the Coulomb interaction.
    To avoid double-counting, it is necessary to define the projector in the $x$ cluster particle-hole basis to the fragment orbitals used to define the cluster as $P^{\textrm{frag},x}$, and construct the local contribution via democratic partitioning, as
    \begin{equation}
    \begin{aligned}
        \Delta\tilde{\mathcal{K}}^{\textrm{cl},x}_{pqrs} = \frac{1}{4} &\left(
            \Delta\mathcal{K}^{\textrm{cl},x}_{tqrs} P^{\textrm{frag},x}_{pt} +
            \Delta\mathcal{K}^{\textrm{cl},x}_{ptrs} P^{\textrm{frag},x}_{qt}\right. \\
            &\left.+
            \Delta\mathcal{K}^{\textrm{cl},x}_{pqts} P^{\textrm{frag},x}_{rt} +
            \Delta\mathcal{K}^{\textrm{cl},x}_{pqrt} P^{\textrm{frag},x}_{st}
        \right) ,
    \end{aligned}
    \end{equation}
    where Einstein summation is assumed. This ensures that the symmetries of the interaction kernel are maintained, while also maintaining the exact large fragment limit.
    The full interaction kernel can obtained by transforming all $\Delta\tilde{\mathcal{K}}^{\textrm{cl},x}$ from the cluster into the basis spanning the whole system and summing all clusters, appropriately recombining it with the original bare Coulomb kernel.
    }
    
    \toadd{Examples of the long-range modifications to the interaction kernel of the full system at convergence are shown in the section~\ref{sec:1dEHM}, demonstrating the ability for the fragment renormalization to even induce attractive components of the effective interaction.}
    This self-consistency is also combined, without double-counting, with the one-particle self-consistency of DMET to match the one-particle density matrix via a fragment-local one-body correlation potential \cite{Knizia2012,Knizia2013,Wouters2016} optimized as a semidefinite program \citep{Wu2020,ocpb:16,scs,diamond2016cvxpy,agrawal2018rewriting}.
    This iterative procedure is continued until convergence of the first two moments of both the fragment dd-response and single-particle response (one-body density matrix), with the interaction kernel and correlation potential no longer changing.

    \toadd{\subsection{Expectation values in EDMET}}

    \toadd{Before summarizing the overall EDMET scheme and practical considerations, we mention the extraction of observables of interest.
    Fragment-local properties can be obtained directly 
    from expectation values over the correlated cluster solution at convergence. In DMET, non-local static expectation values, such as the energy, are obtained via democratic partitioning of the density matrix expressions for each quantity. However, in EDMET, we are also able to define some portion of the non-cluster-local correlated physics, and therefore it is possible to augment the DMET energy expression with an additional contribution arising from coupling to the bosonic bath in the cluster, describing long-range correlated energy contributions. Appendix~\ref{App:Energy} derives this local fragment energy expression in detail, which we present here as
    \begin{equation}
        \begin{aligned}
        E_\text{frag} =& \sum\limits_{p\in \text{frag}} \left( \sum\limits_{q\in\text{clus}} \frac{t_{pq} + \tilde{h}_{pq}}{2} D_{pq}^\text{cl} + \frac{1}{2} \sum\limits_{qrs\in\text{clus}} (pq|rs) P_{qp|sr}^\text{cl,f} \right) \\
        &+ \frac{1}{2}\sum_{\substack{p\in \text{frag}\\ q\in\text{cluster}}} \sum_{n} ({\tilde V}_{pqn} P_{pq,n}^\text{cl,fb} + {\tilde V}_{qpn}P_{qp,n}^\text{cl,fb}), \label{eq:energy}
        \end{aligned}
    \end{equation}
    where we define the one- and two-body fermionic reduced density matrices within the cluster as $D_{pq}^\text{cl} = \langle c_{q}^{\dagger} c_p \rangle_{\text{cl}}$ and $P_{qp|sr}^\text{cl,f} = \langle c_p^{\dagger} c_r^{\dagger} c_s c_q \rangle_{\text{cl}}$ respectively, with the cluster fermion-boson reduced density matrix as $P^\text{cl,fb}_{qp,n} = \langle c_p^{\dagger} c_q a_n \rangle_\textrm{cl}$. The effective one-body interaction within the cluster is defined as
    \begin{equation}
        \tilde{h}_{pq} = t_{pq} + \sum\limits_{rs} \left[(pq|rs) - (ps|rq) \right] D^\text{env}_{rs}.
        \label{eq:effective_onebody}
    \end{equation}
    The first line of Eq.~\ref{eq:energy} is identical to the DMET energy expression \cite{Wouters2016} (which can also be derived from the Migdal--Galitskii formula \cite{Fertitta2019}), while the final term represents the contribution to the correlation energy from the non-local bosonic charge fluctuations between the fragment and environment. This physics is neglected in the DMET energy expression, as there is no consideration of correlations coupling the fragment beyond the lengthscale of the cluster space.}
    
    \toadd{Finally, given that the RPA equations are solved in the full system with the updated interaction kernel and irreducible polarizability from the one- and two-body self-consistency, we can construct fully non-local and dynamic two-body expectation values from this full system RPA with its locally corrected exchange-correlation physics at convergence. This includes quantities such as the dynamic density-density response of Eq.~\ref{eq:rpaddresponse}, or dynamic spin/charge structure factors depending on the symmetry sector probed. We stress again that while these expectation values make the RPA approximation that the ansatz for the excitations only spans single particle-hole excitations, the interaction kernel and single-particle energies of used to construct these excitations are fully screened and renormalized by the self-consistent local correlated physics. Examples of these converged full system RPA spectra are shown in section~\ref{sec:1dEHM}.}
    
    \toadd{\subsection{Overall EDMET Algorithm}}

\begin{algorithm}[H]
  \caption{EDMET algorithmic pseudo-code.}\label{alg:EDMET-loop}
  \footnotesize
  \begin{algorithmic}[1]
    \State Input initial one-body idempotent density matrix over the full system, $\textbf{D}^{ll,(0)}$
    \State Initialize one-body fock matrix, $\hat{f}^{(0)} = \hat{f}(\textbf{D}^{ll,(0)})$, correlation potential $u^{(0)}=0$, correlation kernel $\mathcal{K}^{(0)} = \mathcal{K}_\text{coulomb}$
    \While{correlation potential $u^{(k)}$ and kernel $\mathcal{K}^{(k)}$ have not converged}
    \State Diagonalize $\hat{H}_{ll} = \hat{f}^{(k)} + u^{(k)}$ for $\textbf{D}^{ll,(k)}$
    \If{charge consistent}
    \State Self-consistently solve $\hat{f}^{(k+1)} = \hat{f}(\textbf{D}^{ll,(k)})$
    \Else{}
    \State Set $\hat{f}^{(k+1)} = \hat{f}^{(k)}$
    \EndIf
    \State Perform \ac{RPA} calculation using eigenvalues of $\hat{f}^{(k+1)}$ and $\mathcal{K}^{(k)}$ for $\mat{X}$,$\mat{Y}$ and $\mat{\Omega}$.
    \State Construct electron-boson cluster Hamiltonian $\hat{H}^{\text{emb}}_x$ for all fragments $x$
    \State Set m=0, $\nu^{(0)} = \mu^{(k)}$, to solve for the global chemical potential
    \While{chemical potential $\nu^{(m)}$ is not converged}
        \State Solve $\hat{H}^{\text{emb}}_x - \nu^{(m)} \hat{N}^\text{frag}$ for $\mat{D}^{\text{hl,frag}}_{x}$ for all $x$
        \State Update $\nu^{(m+1)}$ such that Tr($\mat{D}^{\text{hl,frag}}) = N$, the total number of electrons
        \State Set $m \leftarrow m+1$
    \EndWhile
    \State Set $\mu^{(k+1)} = \nu^{(m)}$
    \State Construct $\mat{\chi}_x^{\text{hl,frag}}$, $D_{pq}$, and $P_{pq|rs}$ over the system via democratic partitioning of the cluster RDMs, and $P_{pq,n}^{\textrm{cl,fb}}$ for each cluster.
    \State Compute energy from RDMs.
    \State Update $u^{(k+1)}$ by solving global SDP fitting problem.
    \For{$x$ in 1,...,$N_f$}
    \State Obtain $\mathcal{K}^{(k+1)}_x$ in cluster
    \EndFor
    \State Construct $\mathcal{K}^{(k+1)}$ via democratic partitioning.
    \State Set $k \leftarrow k+1$
    \EndWhile
  \end{algorithmic}
  \label{algo:overall}
\end{algorithm}

\toadd{In Algorithm~\ref{algo:overall}, we sketch pseudocode for the overall EDMET algorithm.} \toadd{Much of the algorithm is the same as a traditional DMET calculations, with changes for the boson bath construction and its cluster hamiltonian via RPA, update of the interaction kernel each iteration, and energy expressions.}
A charge-consistent approach was used throughout this work (where the full system one-body density relaxes due to the correlations), along with \ac{DIIS} acceleration for convergence of both the correlation potential and interaction kernel. 

There are a number of minor modifications possible to the algorithm, depending on whether certain quantities want to be fully converged, or only partly updated each iteration, as well as whether there is only a single fragment space, or if the entire system is partitioned into disjoint fragments. 
In the 1D extended Hubbard model results, it was found that convergence was more stable with the charge self-consistency in the fock matrix as the outer loop, and the correlation kernel convergence as the inner loop, with the density only updated once per (outer-)iteration. Despite these variations, all calculations resulted in tight convergence of the fully self-consistent correlation potential, density and interaction kernel (a fully stationary global mean-field and RPA solution). \toadd{By starting a calculation from a converged interaction kernel and correlation potential of a different hamiltonian, solutions in a particular phase can be continued through the parameter space until they are unstable, thereby also mapping out regions of stable coexistance of different phases.}

\toadd{The EDMET scheme inherits various exactness criteria from the interacting-bath DMET, namely that it is exact for the uncorrelated limit, and the limit where the fragment space spans at least half of the total number of degrees of freedom in the full system. In these limits, the bosonic bath decouples, since there are no correlation-driven charge fluctuations from the resulting cluster into the environment. Furthermore, if coupling to the bosonic space in the correlated cluster model is suppressed in Eq.~\ref{eq:secquantham} (e.g. by restricting the Hilbert space of the solution to have no bosonic occupancy), then the algorithm also rigorously returns to an interacting-bath DMET calculation, including all expectation values, and the overall self-consistency.}
    
\toadd{\section{Results}}

\toadd{\subsection{Extended Hubbard Model} \label{sec:1dEHM}} 

We consider the half-filled extended Hubbard model in both 1D and 2D square lattices, where non-local interactions are introduced (in addition to the local $U$ term) via a nearest-neighbour density-density repulsion with strength $V$. \toadd{This is described by the hamiltonian,
\begin{equation}
    H_\text{EHM} = -t \sum\limits_{\langle ij \rangle \sigma} {\hat c}_{i\sigma}^\dagger {\hat c}_{j\sigma} + U \sum\limits_i {\hat n}_{i\uparrow} {\hat n}_{i\downarrow} + V \sum\limits_{\langle ij\rangle} {\hat n}_i {\hat n}_j,
\end{equation}
    where $\langle ij \rangle$ denotes pairs of nearest neighbouring sites, $\sigma$ denotes spin polarization, and ${\hat n}_{i \sigma}$ is the number operator for spin $\sigma$ on site $i$, where ${\hat n}_i=\sum_{\sigma} {\hat n}_{i \sigma}$. We set $t=1$ in all results to define the energy scale.
}
We first consider a 1D model, to allow comparison to \ac{DMRG} in this low-dimensional limit, as well as other (including embedded) approaches \citep{Sengupta2002,Tsuchiizu2002, Jeckelmann2002,Ejima2007, Lanata2020, Lee2019}.

\begin{figure}
    \centering
    \subfloat{{\includegraphics[width=.47\textwidth]{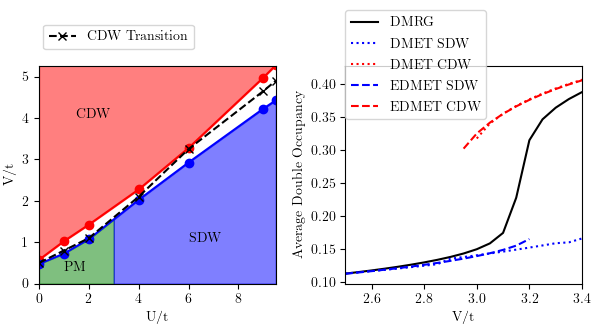} }}%
    \caption{
    Phase diagram of the 1D extended Hubbard model (left), with \toadd{charge-density wave (CDW)}, \toadd{spin-density wave (SDW)} and paramagnetic (PM) \toadd{regions of the parameter space}. The white region indicates a stable coexistence of CDW and SDW phases for this choice of lattice and fragment size, with the dotted line indicating the lowest value of $V/t$ at which CDW becomes energetically favoured, \toadd{indicating the relative energetic stability of the phase}. 
    (right) Double occupancy of \ac{AFM} and \ac{CO} \ac{EDMET} solutions, averaged over the two fragment sites at $U/t=6$, \toadd{showing the coexistance and phase transition behaviour between the regions.}
    }
    \label{fig:1D_EHubbard_order_params}
\end{figure}

The phase diagram of this model is dominated by a \toadd{short-range anti-ferromagnetic spin density wave phase (SDW)} at low $V/U$, and a two-site \toadd{charge density wave phase (CDW)} at high $V/U$ \cite{Ejima2007}. We describe this in EDMET with a two-site fragment, translationally repeated through the lattice, \toadd{which we take to consist of 32 sites with anti-periodic boundary conditions. Spin polarization is allowed to spontaneously break in the static mean-field description of the system.} \vthree{We can further consider the stability of different symmetry solutions by initializing from different symmetry-broken effective correlation potentials in the starting mean-field state.} The correlated cluster each iteration then consists of 4 fermions and 8 bosons. By analyzing the converged \toadd{static} spin and charge densities \vthree{between sublattices}, we can determine the dominant order, regions of coexistance of these phases (where we can stably converge to either phase based on the choice of starting conditions), and the energy of these phases to determine relative stability. These results are summarized in the phase diagram of Fig.~\ref{fig:1D_EHubbard_order_params}, showing the dominant order, regions of phase coexistence, and lowest value of $V/t$ at which symmetry-breaking to a \toadd{charge density wave} phase is energetically favored.
The phase transition line to the CDW state is defined as the lowest value of $V/t$ at which the SDW solution becomes higher in energy than the CDW solution, or destabilises to this CDW solution through the self-consistency. 

\begin{figure}
    \centering
    \includegraphics[width=0.5\textwidth]{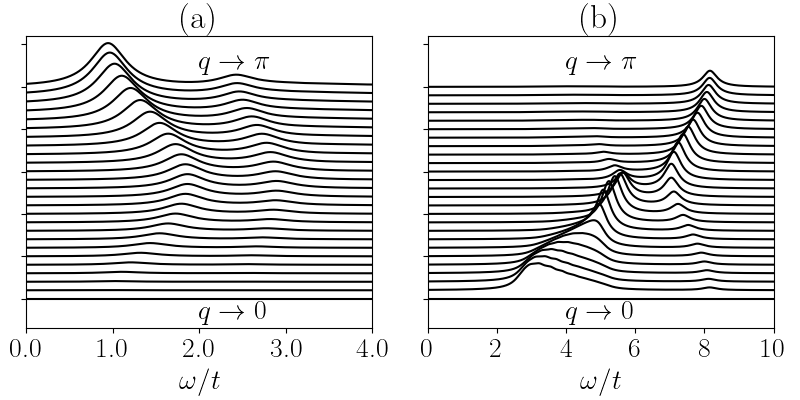}
    \caption{EDMET spectra of (a) the dynamical spin structure factor and (b) the renormalised charge structure factor of the 50-site extended Hubbard model with $U/t=7.8$ and $V=1.3t$.
    A broadening of 0.25t was used.}
    \label{fig:1D_EHubbard_dynamical_structure}
\end{figure}

This phase diagram is \toadd{generally} in good agreement with DMRG and sign-problem-free QMC studies, where these phase diagrams can be found in Refs.~\onlinecite{Sengupta2002,Tsuchiizu2002, Jeckelmann2002,Ejima2007}. However, a very small portion of the true phase diagram \toadd{between the CDW and SDW regions} is also occupied by a \ac{BOW} phase, found in DMRG and QMC studies and characterized by dimerization reflected in the one-body coherences between sites  \citep{Nakamura1999,Sengupta2002,Ejima2007,Liu2011}. While we find a continuous change in this order parameter, the fragmentation of the lattice into two-site fragments biases towards bond order being present even in CDW and SDW phases, \toadd{due to the necessity of the correlation potential to drive the correlated physics in the mean-field picture.} This \ac{BOW} phase is therefore neglected in the phase diagram, \toadd{with its location largely instead taken up by a coexistance region between the two dominant phases.}
\vthree{The phase diagram also shows a small region of paramagnetic (PM) phase at low values of both $U$ and $V$, where the mean-field is neither spin- nor charge-polarised. This is defined by the absence of symmetry-breaking in the single-particle picture either with respect to charge or spin ordering between sites on different sublattices. This phase is not present in more accurate DMRG studies \cite{Ejima2007}, and is likely artificially stabilized due to the small fragment size and the absence of symmetry-breaking in the parent UHF approach at these points. Further studies are required to confirm the rate at which this PM region vanishes as the fragment size increases.}

For a more quantitative comparison to exact \ac{DMRG} \toadd{and \ac{DMET} results}, we also focus on a $U=6t$ cut through the phase diagram, and compare the double occupancy obtained for two-site fragment \ac{EDMET} \toadd{and \ac{DMET}} results as $V/t$ is varied across the transition point, \toadd{characterized by the parameter $\langle {\hat n}_{i \uparrow} {\hat n}_{i \downarrow} \rangle_{\text{frag}}$. The change in this parameter is expected to be discontinuous at the transition (for the infinite system), as the spin-density wave suppresses the local charge fluctuations, resulting in a lower value for this average double occupancy per site in this phase than the CDW.} The discrepancy of this parameter to DMRG is largest about the phase transition point \toadd{for both embedded approaches}, but accurately tracks it outside the vicinity of the transition. \toadd{The predicted transition points also agree well, at $V=3.25t$ and $V=3.2t$ respectively for \ac{EDMET} and \ac{DMET}.} The \ac{EDMET} also predicts a \toadd{narrow} range of phase coexistence between $2.95t \leq V \leq 3.2t$, where the \ac{SDW} state is energetically favored \toadd{till it spontaneous destabilises to the \acs{CDW}. However in comparison, initialising \acs{DMET} (with interacting bath) in a spin-symmetry broken \acs{SDW} phase does not result in collapse to a \acs{CDW} solution, with an unreasonable wide range of stability. While this solution is not competitive with the \acs{CDW} in energy, it remains stable until at least $V>4.4t$, far beyond physical relevance.}


\toadd{A particularly important advantage of the EDMET approach is that it is self-consistent with respect to (limited) dynamical information of two-particle quantities.}
We therefore compute full momentum and energy-resolved two-particle correlation functions via the lattice RPA, where the fragment-local (exchange-)correlation has been self-consistently resolved \toadd{via the EDMET up to first order in the dynamics}. We compute the dynamical spin structure factor, \toadd{defined as
\begin{equation}
    S(q,\omega) = -\pi^{-1} \textrm{Im} \langle 0| \hat{S}_q^{z\dagger} \frac{1}{\hat{H} + \omega - E_0 - i \eta}  \hat{S}_q^{z} |0 \rangle
\end{equation}
and renormalised charge structure factor,
\begin{equation}
    N(q,\omega) = -\pi^{-1}q^{-2} \textrm{Im} \langle 0| \hat{n}_q^\dagger \frac{1}{\hat{H} + \omega - E_0 - i \eta}  \hat{n}_q |0 \rangle,
\end{equation}
where $\hat{S}_q^z$ is the Fourier transform of the local spin density difference, $\hat{S}^z_l=\hat{n}_{l,\uparrow} - \hat{n}_{l,\downarrow}$, and $\hat{n}_l = \hat{n}_{l,\uparrow} + \hat{n}_{l,\downarrow}$ is the local density. These quantities directly characterize the nature of the spin and charge fluctuations at different length and energy scales, and in the \ac{RPA} can be obtained from Eq.~\eqref{eq:rpaddresponse}.}

\begin{figure}[h]
    \centering
    \subfloat{{\includegraphics[width=.4\textwidth]{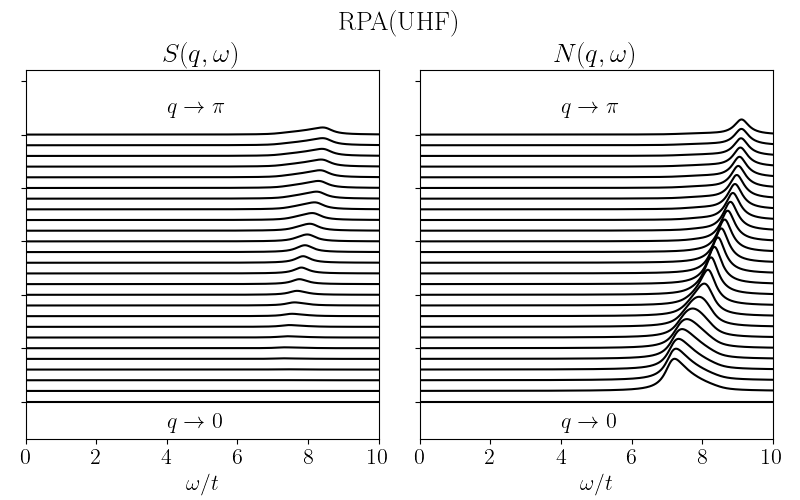} }}%
    \qquad
    \subfloat{{\includegraphics[width=.4\textwidth]{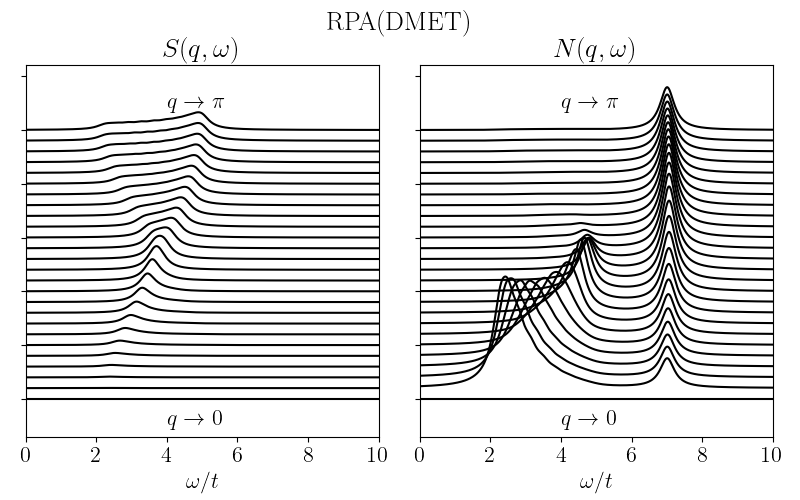} }}%
    \caption{Spectral functions of the dynamical spin structure factor and renormalised charge structure factor for the 50-site extended Hubbard model at half-filling with $U/t=7.8$ and $V=1.3t$ calculated via (a) RPA upon an unrestricted HF calculation, (b) RPA upon a converged DMET mean-field state\toadd{, where the bare Coulomb interaction is used.}
    A finite broadening of 0.25t was used in all cases. Note that the scale is different to Fig.~\ref{fig:1D_EHubbard_dynamical_structure}.}
    \label{fig:1D_EHubbard_structure}
\end{figure}

\toadd{We take a point in the phase diagram at $U/t=7.8$, $V/t=1.3$, which according to optical conductivity measurements is representative of the low-energy physics of $\textrm{SrCuO}_2$ \citep{Kim2004}. The two-site fragment EDMET spin and charge structure factors with a 50 site lattice are presented in}
Fig.~\ref{fig:1D_EHubbard_dynamical_structure}.
The spin structure factor is dominated by a low-energy band which vanishes at $q=0$, with a spin gap at $q=\pi$ and a maximum energy at $q=\pi/2$. Secondary peaks at higher energy define the two-spinon continuum. The charge structure factor is dominated by a broad resonance monotonically decreasing in energy as $q$ decreases. These \toadd{features qualitatively match the} equivalent dynamical DMRG results of Ref.~\onlinecite{Benthien2007}. The main discrepancy from DMRG is the fact that the EDMET spin gap does not strictly vanish (but reaches a minimum) at $q\rightarrow \pi$, while the DMRG spin gap strictly vanishes. This is likely due to the lack of longer-ranged exchange contributions beyond the fragment. \toadd{However, it is somewhat surprising that the EDMET dynamic spin correlation functions are so well reproduced, given that the couplings in the RPA are included between charge fluctuations only, rather than spin fluctuations.}

We can demonstrate the importance of both the strong local correlations, as well as the screening and bosonic bath fluctuations beyond DMET in this cluster model, by comparing these results to both standard \ac{RPA} where the irreducible polarizability is derived from the orbital energies of Hartree--Fock, or the converged \ac{DMET} self-consistent one-body description (decoupling the bosonic bath). \toadd{Both of these use the {\em bare} Coulomb interaction lacking the local two-body correlation found in EDMET. These dynamic correlation functions are presented in Fig.~\ref{fig:1D_EHubbard_structure}, and are qualitatively incorrect, demonstrating} significant redistribution of spectral weight from the inclusion of the bosonic space of EDMET, accounting for the neglect of self-consistent non-local interactions in DMET.

\begin{figure}[h]
    \centering
    \subfloat[\centering U/t = 7.8, V/t = 1.3]{{\includegraphics[width=.4\textwidth]{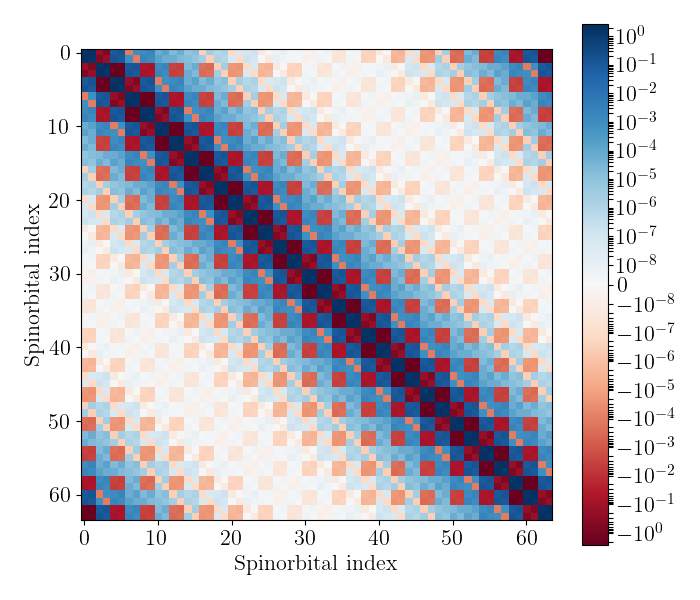} }}%
    \qquad
    \subfloat[\centering U/t = 7.8, V/t = 5.0]{{\includegraphics[width=.4\textwidth]{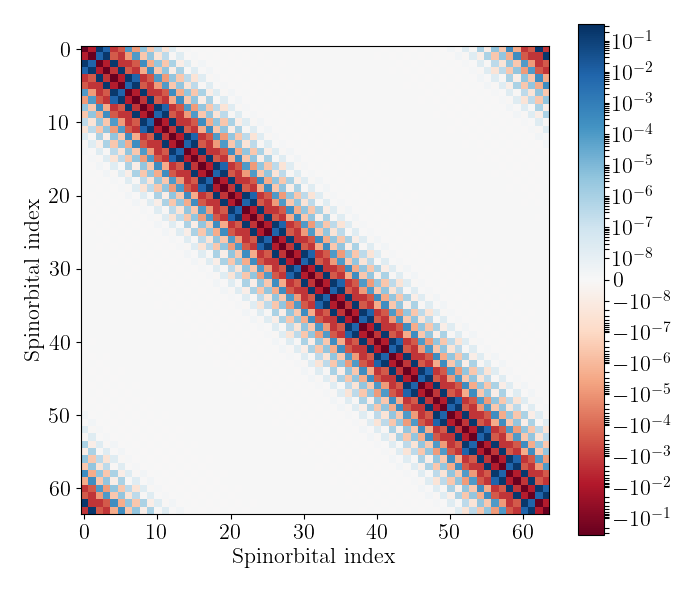} }}%
    \caption{Modifications to the two-point component of the interaction (Coulomb-exchange-correlation) kernel in the lattice of the EDMET due to the self-consistency ($\Delta\tilde{\mathcal{K}}_{ppqq} = \mathcal{K} - {\mat v}$) on a symlog scale. This demonstrates the renormalization of the effective interaction to account for fragment exchange and correlation effects, and is used in the resulting RPA. These are obtained in the half-filled 32-site 1D extended Hubbard model, with (a) $U/t=7.8$ and $V/t=1.3$, or (b) $V/t=5$.
    Even (odd) spin-orbital indices correspond to $\alpha$ ($\beta$) spin states along the chain.}
    \label{fig:1D_EHubbard_RPA_kernel}
\end{figure}

\toadd{Finally, it is instructive to look at the converged effective screened interaction, where these interactions have been modified by the EDMET to match the local dd-response moments. This (static) modification to the long-range Coulomb-exchange-correlation kernel ($\Delta\tilde{\mathcal{K}}_{pqrs} = \mathcal{K} - {\mat v}$) in the SDW and CDW phases is shown in Fig.~\ref{fig:1D_EHubbard_RPA_kernel}, where the local two-point projection of this interaction is depicted, i.e. $\Delta\tilde{\mathcal{K}}_{ppqq}$. It is found that there are attractive lengthscales of this effective interaction between the electrons which are induced by the renormalization due to the strong correlations of the fragment. Translational symmetry of this effective interaction kernel is maintained, but symmetry between the alpha and beta electrons can be broken. It is also found that the lengthscale of the effective screened interaction is longer in the SDW phase than the CDW.}

\begin{figure}
    \centering
    \subfloat{{\includegraphics[width=0.5\textwidth]{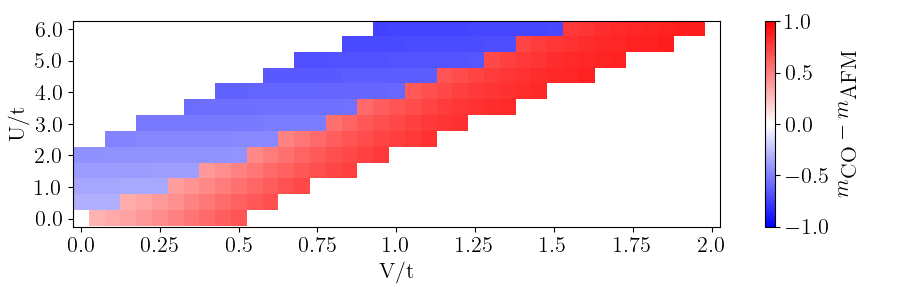} }}%
    \caption{The phase transition in the 6$\times$6 2D extended Hubbard model, showing the charge versus spin order of the lowest energy stable phase, defined over a $2\times2$ plaquette \toadd{built from coupled two-site fragments}. 
    }
    \label{fig:2D_EHubbard_EoS}
\end{figure}

While DMRG provides a benchmark for the extended Hubbard model in 1D, this is no longer the case in 2D. We extend EDMET to the 2D lattice, where symmetries within a $2\times2$ plaquette are characterized via a self-consistently coupled set of two, 2-site fragment clusters. This results in the zero-temperature phase diagram of Fig.~\ref{fig:2D_EHubbard_EoS} showing the magnitude of the staggered magnetization versus charge order.
\toadd{Within our approach these are evaluated from the explicit local one-body symmetry breaking of the mean-field reference state, which at convergence is equal to the local high-level one-body properties.
This gives order parameters
\begin{align}
    m_\textrm{CO} &= \frac{\textrm{Tr}\left(|D^{\textrm{ll},(1)} - D^{\textrm{ll},(2)}|\right)}{\textrm{Tr}\left(D^\textrm{ll}\right)},\\
    m_\textrm{AFM} &= \frac{\textrm{Tr}\left(|D^{\textrm{ll},\alpha} - D^{\textrm{ll},\beta}|\right)}{\textrm{Tr}\left(D^\textrm{ll}\right)}.
\end{align}
Here $D^{(i)}$ and $D^{\sigma}$ are the density matrices on one of the two sublattices defining charge-ordering and the $\sigma$-spin density, respectively.}
\toadd{As with the 1D model, a small parameter region around the phase transition was found to be stable in both phases, depending on the initial conditions for the optimization. In this case, the lowest energy solution between the two phases was chosen in the plot. However, at no point was both charge ordering and staggered magnetization found to co-exist in the same EDMET solution.}

This phase diagram found to qualitatively agree with previous two-particle embedding methods \citep{Onari2004,Ayral2013,VanLoon2014,Medvedeva2017}, though is not directly comparable due to the finite temperature nature of these methods. Specifically, we find no stable paramagnetic phase at zero-temperature, with only checkerboard CDW or SDW phases stable. The lack of PM phase at half filling agrees with other ground state Hubbard model results at $V/t=0$\citep{Zheng2016,Simkovic2020,Wu2020}. 
\toadd{However, for a more quantitative comparison between these EDMET results and state-of-the-art methods, a careful analysis of finite-size scaling of the lattice in the results is needed, as well as a consideration of cluster size convergence 
\cite{Zheng2017b}. This will require methodological extensions of the EDMET to reduce the formal scaling of the current implementation, which will be discussed in Sec.~\ref{sec:concs}.}

\toadd{\subsection{PPP Model}}

The most significant benefits in the formulation are likely to arise in polarizable systems with longer-range interactions, moving towards {\em ab initio} applications \toadd{where e.g. polarons and dispersion physics are found}. In this direction, we consider the \ac{PPP} model, where a more realistic $r^{-1}$-like Coulomb interaction is parameterized to match the low-energy optical properties of poly-acene molecules \citep{Ohno1964,Chandross1997}.
Their low-energy behaviour is determined by a conjugated manifold of polarizable $\pi$-electrons, with a correlated description necessary for accurate spectra. \toadd{This long-range PPP Hamiltonian is parametrised by the Ohno relationship\citep{Ohno1964}, given by
\begin{align}
    H_\text{PPP} &= -t \sum\limits_{\langle ij \rangle \sigma} {\hat c}_{i\sigma}^\dagger {\hat c}_{j\sigma} + U \sum\limits_i {\hat n}_{i\uparrow} {\hat n}_{i\downarrow}\\
    &\hspace{1cm}+ \sum\limits_{i<j} V_{ij} ({\hat n}_{i} - 1)({\hat n}_{j} - 1) \nonumber\\
    V_{ij} &= U / \left(\kappa_{ij} \left(1+ 0.6117 R_{ij}^2\right)^{1/2}\right).
\end{align}
The screened parameterisation of the PPP model (defined in Ref.~\onlinecite{Chandross1997}) defines $U = 8.0$eV, $\kappa_{i,j} = \delta_{i,j} + 2.0 (1-\delta_{i,j})$, $t = 2.4$eV, where $R_{ij}$ is the Euclidean distance between sites $i$ and $j$.}

\begin{figure}
    \centering
    \includegraphics[width=0.49\textwidth]{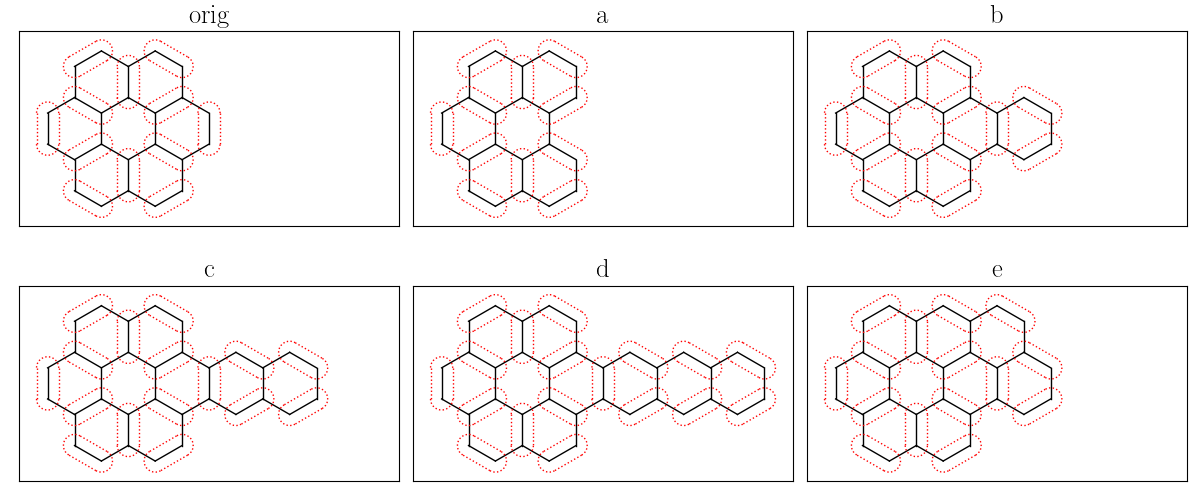}
    \caption{Structures and fragmentations of the coronene derivatives treated in this work.
    Solid lines indicate nearest neighbors, while red dotted lines indicate the specific fragmentation of the lattice chosen.
    All two-site clusters were solved separately and self-consistently coupled, with no symmetries exploited.}
    \label{fig:PPP_fragmentations}
\end{figure}

We apply the EDMET to six different coronene derivatives, considering the resulting optical gap, key to their photo-activity. High-level multi-reference \toadd{configuration interaction with singles and doubles (MRCISD) optical gaps are available for comparison from Ref.~\cite{Bhattacharyya2020}, and are expected to be highly accurate benchmarks.} The tight-binding lattices corresponding to the atomic configurations are partitioned into self-consistently coupled two-site fragments. \toadd{These different coronene derivatives, along with the specific choice of partitioning into two-atom fragments is shown in Fig.~\ref{fig:PPP_fragmentations}, corresponding to the models in Ref.~\onlinecite{Bhattacharyya2020}.}

\begin{table}
    \begin{tabular}{|c|c|c|c|c|c|c|c|}
        \hline
         &\multirow{2}{*}{Method}& \multicolumn{6}{c|}{Coronene Derivative}\\
        \cline{3-8}
         & & orig & a & b & c & d & e \\
         \hline
         \multirow{5}{*}{\rotatebox[origin=c]{90}{Gap/eV}}& RPA & 5.08 & 3.85 & 3.95 & 3.57 & 3.14 & 3.43 \\
         & $\text{MRCISD}$ & 4.20 & 3.39 & 3.42 & 2.93 & 2.70 & 2.95 \\
         & EDMET & 4.12 & 3.17 & 3.31 & 3.05 & 2.79 & 2.87 \\
         & $\text{Experiment}$ & 3.55 (4.1) & - & 3.38 & - & - & 3.02 \\
         \hline
    \end{tabular}
\caption{Optical gaps for PPP models of Coronene derivatives.
In the case of coronene (denoted `orig'), the lowest-energy absorption at 3.55eV is dipole forbidden, with the symmetry-allowed gap at 4.1eV provided.
MRCISD and experimental results are taken from Refs.~\onlinecite{Bhattacharyya2020}~and~\onlinecite{Bagley2013} respectively.
}
\label{tab:PPP_opt_gap}
\end{table}

The EDMET results of Table~\ref{tab:PPP_opt_gap} show that the inclusion of local correlation and exchange physics of these fragments results in a significant improvement of the optical gaps across all systems compared to RPA. This reduces the mean relative errors from 17.4\% (at RPA level) to 3.6\% (at EDMET level) compared to MRCISD.
Moreover, the average deviation from experimental results (where available) is 2.0\% and 2.5\% for MRCISD and \ac{EDMET} respectively, suggesting that error in the \ac{EDMET} is likely lower than that inherent in the \ac{PPP} model approximations.

\toadd{\section{Conclusions and outlook} \label{sec:concs}}

\toadd{We have presented a static two-body quantum embedding, EDMET,  which enables incorporation of self-consistent long-range interactions and screening effects between a local fragment and its environment at zero temperature, without requiring explicit frequency dependence in the effective interaction. We have developed an exact mapping from an extended system at the level of RPA, to a cluster model defined by a local fragment, fermionic and bosonic bath spaces.
Construction of this cluster hamiltonian and two-particle self-consistency are both performed algebraically, avoiding any numerical fits, with a fixed upper bound on the number of required bosonic and fermionic bath states, which is independent of the size of the full system. 
It has been shown that the approach can correctly predict the behaviour of long-range interacting lattices in both 1 and 2 dimensions, quantitatively describing phase transitions and static expectations such as double occupancy, as well as dynamical two-particle spectra.}

\toadd{However, the computational costs of the EDMET scheme are more significant than the one-particle DMET embedding scheme on which it builds. These additional costs can be broken down into two parts. First, the cluster mapping and construction of the bosonic bath coupling elements currently requires the full solution to the RPA excitations each iteration, which scales as $\mathcal{O}[N^6]$ -- substantially more than mean-field scaling, and stymieing application to large systems in the thermodynamic limit. Current work is ongoing to reduce this scaling via resolution of the identity approaches, in order to return this to a more manageable $\mathcal{O}[N^4]$ scaling, in keeping with Hatree--Fock theory \citep{Eshuis2010}. Secondly, the solution of the resulting cluster is more computationally demanding for the same fragment size when compared to DMET, due to the additional complexity arising from the coupled bosonic degrees of freedom. We are also therefore exploring other solvers, which will allow the approximate solutions for larger fragments than currently admitted via exact diagonalization \cite{Jeckelmann1998,White2020,Mordovina2020}, as well as approaches to reduce the number of coupled bosons in the cluster via further renormalization of their effect. At that point, we should have further clarity on the advantages of the EDMET approach and value in describing this long-range physics rather than capturing it via explicit enlargement of the fragment or interacting fermionic bath size (as performed in Ref.~\cite{nusspickel2021}). Future work will also enable the approach to be extended naturally to describe coupling to physical bosonic degrees of freedom, such as phonons in the lattice \citep{Sandhoefer2016,Reinhard2019}.}

\section*{Acknowledgements}
G.H.B would like to thank Philipp Werner and Garnet Chan for helpful discussions about this work.
G.H.B. also gratefully acknowledges funding from the Royal Society via a University Research Fellowship, as well as funding from the European Research Council (ERC) under the European Union’s Horizon 2020 research and innovation programme (grant agreement No. 759063).


\begin{thebibliography}{94}%
\makeatletter
\providecommand \@ifxundefined [1]{%
 \@ifx{#1\undefined}
}%
\providecommand \@ifnum [1]{%
 \ifnum #1\expandafter \@firstoftwo
 \else \expandafter \@secondoftwo
 \fi
}%
\providecommand \@ifx [1]{%
 \ifx #1\expandafter \@firstoftwo
 \else \expandafter \@secondoftwo
 \fi
}%
\providecommand \natexlab [1]{#1}%
\providecommand \enquote  [1]{``#1''}%
\providecommand \bibnamefont  [1]{#1}%
\providecommand \bibfnamefont [1]{#1}%
\providecommand \citenamefont [1]{#1}%
\providecommand \href@noop [0]{\@secondoftwo}%
\providecommand \href [0]{\begingroup \@sanitize@url \@href}%
\providecommand \@href[1]{\@@startlink{#1}\@@href}%
\providecommand \@@href[1]{\endgroup#1\@@endlink}%
\providecommand \@sanitize@url [0]{\catcode `\\12\catcode `\$12\catcode
  `\&12\catcode `\#12\catcode `\^12\catcode `\_12\catcode `\%12\relax}%
\providecommand \@@startlink[1]{}%
\providecommand \@@endlink[0]{}%
\providecommand \url  [0]{\begingroup\@sanitize@url \@url }%
\providecommand \@url [1]{\endgroup\@href {#1}{\urlprefix }}%
\providecommand \urlprefix  [0]{URL }%
\providecommand \Eprint [0]{\href }%
\providecommand \doibase [0]{https://doi.org/}%
\providecommand \selectlanguage [0]{\@gobble}%
\providecommand \bibinfo  [0]{\@secondoftwo}%
\providecommand \bibfield  [0]{\@secondoftwo}%
\providecommand \translation [1]{[#1]}%
\providecommand \BibitemOpen [0]{}%
\providecommand \bibitemStop [0]{}%
\providecommand \bibitemNoStop [0]{.\EOS\space}%
\providecommand \EOS [0]{\spacefactor3000\relax}%
\providecommand \BibitemShut  [1]{\csname bibitem#1\endcsname}%
\let\auto@bib@innerbib\@empty
\bibitem [{\citenamefont {Kent}\ and\ \citenamefont
  {Kotliar}(2018)}]{Kent2018}%
  \BibitemOpen
  \bibfield  {author} {\bibinfo {author} {\bibfnamefont {P.~R.}\ \bibnamefont
  {Kent}}\ and\ \bibinfo {author} {\bibfnamefont {G.}~\bibnamefont {Kotliar}},\
  }\href@noop {} {\bibfield  {journal} {\bibinfo  {journal} {Science (80-. ).}\
  }\textbf {\bibinfo {volume} {361}},\ \bibinfo {pages} {348} (\bibinfo {year}
  {2018})}\BibitemShut {NoStop}%
\bibitem [{\citenamefont {Wagner}\ and\ \citenamefont
  {Abbamonte}(2014)}]{Wagner2014}%
  \BibitemOpen
  \bibfield  {author} {\bibinfo {author} {\bibfnamefont {L.~K.}\ \bibnamefont
  {Wagner}}\ and\ \bibinfo {author} {\bibfnamefont {P.}~\bibnamefont
  {Abbamonte}},\ }\href@noop {} {\bibfield  {journal} {\bibinfo  {journal}
  {Phys. Rev. B - Condens. Matter Mater. Phys.}\ }\textbf {\bibinfo {volume}
  {90}},\ \bibinfo {pages} {125129} (\bibinfo {year} {2014})}\BibitemShut
  {NoStop}%
\bibitem [{\citenamefont {Alling}\ \emph {et~al.}(2010)\citenamefont {Alling},
  \citenamefont {Marten},\ and\ \citenamefont {Abrikosov}}]{Alling2010}%
  \BibitemOpen
  \bibfield  {author} {\bibinfo {author} {\bibfnamefont {B.}~\bibnamefont
  {Alling}}, \bibinfo {author} {\bibfnamefont {T.}~\bibnamefont {Marten}},\
  and\ \bibinfo {author} {\bibfnamefont {I.~A.}\ \bibnamefont {Abrikosov}},\
  }\href {https://journals.aps.org/prb/abstract/10.1103/PhysRevB.82.184430}
  {\bibfield  {journal} {\bibinfo  {journal} {Phys. Rev. B - Condens. Matter
  Mater. Phys.}\ }\textbf {\bibinfo {volume} {82}},\ \bibinfo {pages} {184430}
  (\bibinfo {year} {2010})}\BibitemShut {NoStop}%
\bibitem [{\citenamefont {Basov}\ \emph {et~al.}(2011)\citenamefont {Basov},
  \citenamefont {Averitt}, \citenamefont {{Van Der Marel}}, \citenamefont
  {Dressel},\ and\ \citenamefont {Haule}}]{Basov2011}%
  \BibitemOpen
  \bibfield  {author} {\bibinfo {author} {\bibfnamefont {D.~N.}\ \bibnamefont
  {Basov}}, \bibinfo {author} {\bibfnamefont {R.~D.}\ \bibnamefont {Averitt}},
  \bibinfo {author} {\bibfnamefont {D.}~\bibnamefont {{Van Der Marel}}},
  \bibinfo {author} {\bibfnamefont {M.}~\bibnamefont {Dressel}},\ and\ \bibinfo
  {author} {\bibfnamefont {K.}~\bibnamefont {Haule}},\ }\href@noop {}
  {\bibfield  {journal} {\bibinfo  {journal} {Rev. Mod. Phys.}\ }\textbf
  {\bibinfo {volume} {83}},\ \bibinfo {pages} {471} (\bibinfo {year}
  {2011})}\BibitemShut {NoStop}%
\bibitem [{\citenamefont {Boehnke}\ \emph {et~al.}(2016)\citenamefont
  {Boehnke}, \citenamefont {Nilsson}, \citenamefont {Aryasetiawan},\ and\
  \citenamefont {Werner}}]{Boehnke2016}%
  \BibitemOpen
  \bibfield  {author} {\bibinfo {author} {\bibfnamefont {L.}~\bibnamefont
  {Boehnke}}, \bibinfo {author} {\bibfnamefont {F.}~\bibnamefont {Nilsson}},
  \bibinfo {author} {\bibfnamefont {F.}~\bibnamefont {Aryasetiawan}},\ and\
  \bibinfo {author} {\bibfnamefont {P.}~\bibnamefont {Werner}},\ }\href@noop {}
  {\bibfield  {journal} {\bibinfo  {journal} {Phys. Rev. B}\ }\textbf {\bibinfo
  {volume} {94}},\ \bibinfo {pages} {201106} (\bibinfo {year}
  {2016})}\BibitemShut {NoStop}%
\bibitem [{\citenamefont {Sun}\ and\ \citenamefont {Chan}(2016)}]{Sun2016}%
  \BibitemOpen
  \bibfield  {author} {\bibinfo {author} {\bibfnamefont {Q.}~\bibnamefont
  {Sun}}\ and\ \bibinfo {author} {\bibfnamefont {G.~K.~L.}\ \bibnamefont
  {Chan}},\ }\href@noop {} {\bibfield  {journal} {\bibinfo  {journal} {Acc.
  Chem. Res.}\ }\textbf {\bibinfo {volume} {49}},\ \bibinfo {pages} {2705}
  (\bibinfo {year} {2016})}\BibitemShut {NoStop}%
\bibitem [{\citenamefont {Georges}\ \emph {et~al.}(1996)\citenamefont
  {Georges}, \citenamefont {Kotliar}, \citenamefont {Krauth},\ and\
  \citenamefont {Rozenberg}}]{Georges1996}%
  \BibitemOpen
  \bibfield  {author} {\bibinfo {author} {\bibfnamefont {A.}~\bibnamefont
  {Georges}}, \bibinfo {author} {\bibfnamefont {G.}~\bibnamefont {Kotliar}},
  \bibinfo {author} {\bibfnamefont {W.}~\bibnamefont {Krauth}},\ and\ \bibinfo
  {author} {\bibfnamefont {M.~J.}\ \bibnamefont {Rozenberg}},\ }\href@noop {}
  {\bibfield  {journal} {\bibinfo  {journal} {Rev. Mod. Phys.}\ }\textbf
  {\bibinfo {volume} {68}},\ \bibinfo {pages} {13} (\bibinfo {year}
  {1996})}\BibitemShut {NoStop}%
\bibitem [{\citenamefont {Hirsch}\ and\ \citenamefont
  {Fye}(1986)}]{Hirsch1986}%
  \BibitemOpen
  \bibfield  {author} {\bibinfo {author} {\bibfnamefont {J.~E.}\ \bibnamefont
  {Hirsch}}\ and\ \bibinfo {author} {\bibfnamefont {R.~M.}\ \bibnamefont
  {Fye}},\ }\href@noop {} {\bibfield  {journal} {\bibinfo  {journal} {Phys.
  Rev. Lett.}\ }\textbf {\bibinfo {volume} {56}},\ \bibinfo {pages} {2521}
  (\bibinfo {year} {1986})}\BibitemShut {NoStop}%
\bibitem [{\citenamefont {Werner}\ \emph {et~al.}(2006)\citenamefont {Werner},
  \citenamefont {Comanac}, \citenamefont {{De Medici}}, \citenamefont
  {Troyer},\ and\ \citenamefont {Millis}}]{Werner2006}%
  \BibitemOpen
  \bibfield  {author} {\bibinfo {author} {\bibfnamefont {P.}~\bibnamefont
  {Werner}}, \bibinfo {author} {\bibfnamefont {A.}~\bibnamefont {Comanac}},
  \bibinfo {author} {\bibfnamefont {L.}~\bibnamefont {{De Medici}}}, \bibinfo
  {author} {\bibfnamefont {M.}~\bibnamefont {Troyer}},\ and\ \bibinfo {author}
  {\bibfnamefont {A.~J.}\ \bibnamefont {Millis}},\ }\href@noop {} {\bibfield
  {journal} {\bibinfo  {journal} {Phys. Rev. Lett.}\ }\textbf {\bibinfo
  {volume} {97}} (\bibinfo {year} {2006})}\BibitemShut {NoStop}%
\bibitem [{\citenamefont {Werner}\ and\ \citenamefont
  {Millis}(2006)}]{Werner2006a}%
  \BibitemOpen
  \bibfield  {author} {\bibinfo {author} {\bibfnamefont {P.}~\bibnamefont
  {Werner}}\ and\ \bibinfo {author} {\bibfnamefont {A.~J.}\ \bibnamefont
  {Millis}},\ }\href@noop {} {\bibfield  {journal} {\bibinfo  {journal} {Phys.
  Rev. B - Condens. Matter Mater. Phys.}\ }\textbf {\bibinfo {volume} {74}}
  (\bibinfo {year} {2006})}\BibitemShut {NoStop}%
\bibitem [{\citenamefont {Zgid}\ \emph {et~al.}(2012)\citenamefont {Zgid},
  \citenamefont {Gull},\ and\ \citenamefont {Chan}}]{Zgid2012}%
  \BibitemOpen
  \bibfield  {author} {\bibinfo {author} {\bibfnamefont {D.}~\bibnamefont
  {Zgid}}, \bibinfo {author} {\bibfnamefont {E.}~\bibnamefont {Gull}},\ and\
  \bibinfo {author} {\bibfnamefont {G.~K.~L.}\ \bibnamefont {Chan}},\
  }\href@noop {} {\bibfield  {journal} {\bibinfo  {journal} {Phys. Rev. B -
  Condens. Matter Mater. Phys.}\ }\textbf {\bibinfo {volume} {86}},\ \bibinfo
  {pages} {165128} (\bibinfo {year} {2012})}\BibitemShut {NoStop}%
\bibitem [{\citenamefont {Lu}\ \emph {et~al.}(2014)\citenamefont {Lu},
  \citenamefont {H{\"{o}}ppner}, \citenamefont {Gunnarsson},\ and\
  \citenamefont {Haverkort}}]{Lu2014}%
  \BibitemOpen
  \bibfield  {author} {\bibinfo {author} {\bibfnamefont {Y.}~\bibnamefont
  {Lu}}, \bibinfo {author} {\bibfnamefont {M.}~\bibnamefont {H{\"{o}}ppner}},
  \bibinfo {author} {\bibfnamefont {O.}~\bibnamefont {Gunnarsson}},\ and\
  \bibinfo {author} {\bibfnamefont {M.~W.}\ \bibnamefont {Haverkort}},\
  }\href@noop {} {\bibfield  {journal} {\bibinfo  {journal} {Phys. Rev. B -
  Condens. Matter Mater. Phys.}\ }\textbf {\bibinfo {volume} {90}},\ \bibinfo
  {pages} {085102} (\bibinfo {year} {2014})}\BibitemShut {NoStop}%
\bibitem [{\citenamefont {Go}\ and\ \citenamefont {Millis}(2017)}]{Go2017}%
  \BibitemOpen
  \bibfield  {author} {\bibinfo {author} {\bibfnamefont {A.}~\bibnamefont
  {Go}}\ and\ \bibinfo {author} {\bibfnamefont {A.~J.}\ \bibnamefont
  {Millis}},\ }\href@noop {} {\bibfield  {journal} {\bibinfo  {journal} {Phys.
  Rev. B}\ }\textbf {\bibinfo {volume} {96}},\ \bibinfo {pages} {085139}
  (\bibinfo {year} {2017})}\BibitemShut {NoStop}%
\bibitem [{\citenamefont {Zhu}\ \emph {et~al.}(2019)\citenamefont {Zhu},
  \citenamefont {Jim{\'{e}}nez-Hoyos}, \citenamefont {McClain}, \citenamefont
  {Berkelbach},\ and\ \citenamefont {Chan}}]{Zhu2019}%
  \BibitemOpen
  \bibfield  {author} {\bibinfo {author} {\bibfnamefont {T.}~\bibnamefont
  {Zhu}}, \bibinfo {author} {\bibfnamefont {C.~A.}\ \bibnamefont
  {Jim{\'{e}}nez-Hoyos}}, \bibinfo {author} {\bibfnamefont {J.}~\bibnamefont
  {McClain}}, \bibinfo {author} {\bibfnamefont {T.~C.}\ \bibnamefont
  {Berkelbach}},\ and\ \bibinfo {author} {\bibfnamefont {G.~K.~L.}\
  \bibnamefont {Chan}},\ }\href@noop {} {\bibfield  {journal} {\bibinfo
  {journal} {Phys. Rev. B}\ }\textbf {\bibinfo {volume} {100}},\ \bibinfo
  {pages} {115154} (\bibinfo {year} {2019})}\BibitemShut {NoStop}%
\bibitem [{\citenamefont {Savrasov}\ \emph {et~al.}(2001)\citenamefont
  {Savrasov}, \citenamefont {Kotliar},\ and\ \citenamefont
  {Abrahams}}]{Savrasov2001}%
  \BibitemOpen
  \bibfield  {author} {\bibinfo {author} {\bibfnamefont {S.~Y.}\ \bibnamefont
  {Savrasov}}, \bibinfo {author} {\bibfnamefont {G.}~\bibnamefont {Kotliar}},\
  and\ \bibinfo {author} {\bibfnamefont {E.}~\bibnamefont {Abrahams}},\
  }\href@noop {} {\bibfield  {journal} {\bibinfo  {journal} {Nature}\ }\textbf
  {\bibinfo {volume} {410}},\ \bibinfo {pages} {793} (\bibinfo {year}
  {2001})}\BibitemShut {NoStop}%
\bibitem [{\citenamefont {Gunnarsson}\ \emph {et~al.}(1989)\citenamefont
  {Gunnarsson}, \citenamefont {Andersen}, \citenamefont {Jepsen},\ and\
  \citenamefont {Zaanen}}]{Gunnarsson1989}%
  \BibitemOpen
  \bibfield  {author} {\bibinfo {author} {\bibfnamefont {O.}~\bibnamefont
  {Gunnarsson}}, \bibinfo {author} {\bibfnamefont {O.~K.}\ \bibnamefont
  {Andersen}}, \bibinfo {author} {\bibfnamefont {O.}~\bibnamefont {Jepsen}},\
  and\ \bibinfo {author} {\bibfnamefont {J.}~\bibnamefont {Zaanen}},\
  }\href@noop {} {\bibfield  {journal} {\bibinfo  {journal} {Phys. Rev. B}\
  }\textbf {\bibinfo {volume} {39}},\ \bibinfo {pages} {1708} (\bibinfo {year}
  {1989})}\BibitemShut {NoStop}%
\bibitem [{\citenamefont {Gunnarsson}(1990)}]{Gunnarsson1990}%
  \BibitemOpen
  \bibfield  {author} {\bibinfo {author} {\bibfnamefont {O.}~\bibnamefont
  {Gunnarsson}},\ }\href@noop {} {\bibfield  {journal} {\bibinfo  {journal}
  {Phys. Rev. B}\ }\textbf {\bibinfo {volume} {41}},\ \bibinfo {pages} {514}
  (\bibinfo {year} {1990})}\BibitemShut {NoStop}%
\bibitem [{\citenamefont {Kotani}(2000)}]{Kotani2000}%
  \BibitemOpen
  \bibfield  {author} {\bibinfo {author} {\bibfnamefont {T.}~\bibnamefont
  {Kotani}},\ }\href@noop {} {\bibfield  {journal} {\bibinfo  {journal} {J.
  Phys. Condens. Matter}\ }\textbf {\bibinfo {volume} {12}},\ \bibinfo {pages}
  {2413} (\bibinfo {year} {2000})}\BibitemShut {NoStop}%
\bibitem [{\citenamefont {Aryasetiawan}\ \emph {et~al.}(2004)\citenamefont
  {Aryasetiawan}, \citenamefont {Imada}, \citenamefont {Georges}, \citenamefont
  {Kotliar}, \citenamefont {Biermann},\ and\ \citenamefont
  {Lichtenstein}}]{Aryasetiawan2004}%
  \BibitemOpen
  \bibfield  {author} {\bibinfo {author} {\bibfnamefont {F.}~\bibnamefont
  {Aryasetiawan}}, \bibinfo {author} {\bibfnamefont {M.}~\bibnamefont {Imada}},
  \bibinfo {author} {\bibfnamefont {A.}~\bibnamefont {Georges}}, \bibinfo
  {author} {\bibfnamefont {G.}~\bibnamefont {Kotliar}}, \bibinfo {author}
  {\bibfnamefont {S.}~\bibnamefont {Biermann}},\ and\ \bibinfo {author}
  {\bibfnamefont {A.~I.}\ \bibnamefont {Lichtenstein}},\ }\href@noop {}
  {\bibfield  {journal} {\bibinfo  {journal} {Phys. Rev. B Condens. Matter}\
  }\textbf {\bibinfo {volume} {70}},\ \bibinfo {pages} {1} (\bibinfo {year}
  {2004})}\BibitemShut {NoStop}%
\bibitem [{\citenamefont {Zhu}\ and\ \citenamefont {Chan}(2021)}]{Zhu2021}%
  \BibitemOpen
  \bibfield  {author} {\bibinfo {author} {\bibfnamefont {T.}~\bibnamefont
  {Zhu}}\ and\ \bibinfo {author} {\bibfnamefont {G.~K.-L.}\ \bibnamefont
  {Chan}},\ }\href@noop {} {\bibfield  {journal} {\bibinfo  {journal} {Phys.
  Rev. X}\ }\textbf {\bibinfo {volume} {11}},\ \bibinfo {pages} {021006}
  (\bibinfo {year} {2021})}\BibitemShut {NoStop}%
\bibitem [{\citenamefont {Almbladh}(1999)}]{Almbladh1999}%
  \BibitemOpen
  \bibfield  {author} {\bibinfo {author} {\bibfnamefont {C.~O.}\ \bibnamefont
  {Almbladh}},\ }\href@noop {} {\bibfield  {journal} {\bibinfo  {journal}
  {International Journal of Modern Physics B}\ }\textbf {\bibinfo {volume}
  {13}},\ \bibinfo {pages} {535} (\bibinfo {year} {1999})}\BibitemShut
  {NoStop}%
\bibitem [{\citenamefont {Biermann}\ \emph {et~al.}(2003)\citenamefont
  {Biermann}, \citenamefont {Aryasetiawan},\ and\ \citenamefont
  {Georges}}]{Biermann2002}%
  \BibitemOpen
  \bibfield  {author} {\bibinfo {author} {\bibfnamefont {S.}~\bibnamefont
  {Biermann}}, \bibinfo {author} {\bibfnamefont {F.}~\bibnamefont
  {Aryasetiawan}},\ and\ \bibinfo {author} {\bibfnamefont {A.}~\bibnamefont
  {Georges}},\ }\href@noop {} {\bibfield  {journal} {\bibinfo  {journal} {Phys.
  Rev. Lett.}\ }\textbf {\bibinfo {volume} {90}},\ \bibinfo {pages} {4}
  (\bibinfo {year} {2003})}\BibitemShut {NoStop}%
\bibitem [{\citenamefont {Haule}\ and\ \citenamefont
  {Kotliar}(2007)}]{Haule2007}%
  \BibitemOpen
  \bibfield  {author} {\bibinfo {author} {\bibfnamefont {K.}~\bibnamefont
  {Haule}}\ and\ \bibinfo {author} {\bibfnamefont {G.}~\bibnamefont
  {Kotliar}},\ }\href@noop {} {\bibfield  {journal} {\bibinfo  {journal} {Phys.
  Rev. B Condens. Matter}\ }\textbf {\bibinfo {volume} {76}} (\bibinfo {year}
  {2007})}\BibitemShut {NoStop}%
\bibitem [{\citenamefont {Ayral}\ \emph {et~al.}(2013)\citenamefont {Ayral},
  \citenamefont {Biermann},\ and\ \citenamefont {Werner}}]{Ayral2013}%
  \BibitemOpen
  \bibfield  {author} {\bibinfo {author} {\bibfnamefont {T.}~\bibnamefont
  {Ayral}}, \bibinfo {author} {\bibfnamefont {S.}~\bibnamefont {Biermann}},\
  and\ \bibinfo {author} {\bibfnamefont {P.}~\bibnamefont {Werner}},\
  }\href@noop {} {\bibfield  {journal} {\bibinfo  {journal} {Phys. Rev. B
  Condens. Matter}\ }\textbf {\bibinfo {volume} {87}},\ \bibinfo {pages}
  {125149} (\bibinfo {year} {2013})}\BibitemShut {NoStop}%
\bibitem [{\citenamefont {Choi}\ \emph {et~al.}(2016)\citenamefont {Choi},
  \citenamefont {Kutepov}, \citenamefont {Haule}, \citenamefont {{Van
  Schilfgaarde}},\ and\ \citenamefont {Kotliar}}]{Choi2016}%
  \BibitemOpen
  \bibfield  {author} {\bibinfo {author} {\bibfnamefont {S.}~\bibnamefont
  {Choi}}, \bibinfo {author} {\bibfnamefont {A.}~\bibnamefont {Kutepov}},
  \bibinfo {author} {\bibfnamefont {K.}~\bibnamefont {Haule}}, \bibinfo
  {author} {\bibfnamefont {M.}~\bibnamefont {{Van Schilfgaarde}}},\ and\
  \bibinfo {author} {\bibfnamefont {G.}~\bibnamefont {Kotliar}},\ }\href@noop
  {} {\bibfield  {journal} {\bibinfo  {journal} {npj Quantum Materials}\
  }\textbf {\bibinfo {volume} {1}} (\bibinfo {year} {2016})}\BibitemShut
  {NoStop}%
\bibitem [{\citenamefont {Lechermann}\ \emph {et~al.}(2017)\citenamefont
  {Lechermann}, \citenamefont {Lichtenstein},\ and\ \citenamefont
  {Potthoff}}]{Lechermann2017}%
  \BibitemOpen
  \bibfield  {author} {\bibinfo {author} {\bibfnamefont {F.}~\bibnamefont
  {Lechermann}}, \bibinfo {author} {\bibfnamefont {A.~I.}\ \bibnamefont
  {Lichtenstein}},\ and\ \bibinfo {author} {\bibfnamefont {M.}~\bibnamefont
  {Potthoff}},\ }\href@noop {} {\bibfield  {journal} {\bibinfo  {journal} {Eur.
  Phys. J. Spec. Top.}\ }\textbf {\bibinfo {volume} {226}},\ \bibinfo {pages}
  {2591} (\bibinfo {year} {2017})}\BibitemShut {NoStop}%
\bibitem [{\citenamefont {Medvedeva}\ \emph {et~al.}(2017)\citenamefont
  {Medvedeva}, \citenamefont {Iskakov}, \citenamefont {Krien}, \citenamefont
  {Mazurenko},\ and\ \citenamefont {Lichtenstein}}]{Medvedeva2017}%
  \BibitemOpen
  \bibfield  {author} {\bibinfo {author} {\bibfnamefont {D.}~\bibnamefont
  {Medvedeva}}, \bibinfo {author} {\bibfnamefont {S.}~\bibnamefont {Iskakov}},
  \bibinfo {author} {\bibfnamefont {F.}~\bibnamefont {Krien}}, \bibinfo
  {author} {\bibfnamefont {V.~V.}\ \bibnamefont {Mazurenko}},\ and\ \bibinfo
  {author} {\bibfnamefont {A.~I.}\ \bibnamefont {Lichtenstein}},\ }\href@noop
  {} {\bibfield  {journal} {\bibinfo  {journal} {Phys. Rev. B}\ }\textbf
  {\bibinfo {volume} {96}},\ \bibinfo {pages} {235149} (\bibinfo {year}
  {2017})}\BibitemShut {NoStop}%
\bibitem [{\citenamefont {Tomczak}\ \emph {et~al.}(2017)\citenamefont
  {Tomczak}, \citenamefont {Liu}, \citenamefont {Toschi}, \citenamefont
  {Kresse},\ and\ \citenamefont {Held}}]{Tomczak2017}%
  \BibitemOpen
  \bibfield  {author} {\bibinfo {author} {\bibfnamefont {J.~M.}\ \bibnamefont
  {Tomczak}}, \bibinfo {author} {\bibfnamefont {P.}~\bibnamefont {Liu}},
  \bibinfo {author} {\bibfnamefont {A.}~\bibnamefont {Toschi}}, \bibinfo
  {author} {\bibfnamefont {G.}~\bibnamefont {Kresse}},\ and\ \bibinfo {author}
  {\bibfnamefont {K.}~\bibnamefont {Held}},\ }\href@noop {} {\bibfield
  {journal} {\bibinfo  {journal} {Eur. Phys. J. Spec. Top.}\ }\textbf {\bibinfo
  {volume} {226}},\ \bibinfo {pages} {2565} (\bibinfo {year}
  {2017})}\BibitemShut {NoStop}%
\bibitem [{\citenamefont {Van~Loon}\ and\ \citenamefont
  {Katsnelson}(2018)}]{VanLoon2018}%
  \BibitemOpen
  \bibfield  {author} {\bibinfo {author} {\bibfnamefont {E.}~\bibnamefont
  {Van~Loon}}\ and\ \bibinfo {author} {\bibfnamefont {M.}~\bibnamefont
  {Katsnelson}},\ }in\ \href@noop {} {\emph {\bibinfo {booktitle} {Journal of
  Physics: Conference Series}}},\ Vol.\ \bibinfo {volume} {1136}\ (\bibinfo
  {organization} {IOP Publishing},\ \bibinfo {year} {2018})\ p.\ \bibinfo
  {pages} {012006}\BibitemShut {NoStop}%
\bibitem [{\citenamefont {Rohringer}\ \emph {et~al.}(2018)\citenamefont
  {Rohringer}, \citenamefont {Hafermann}, \citenamefont {Toschi}, \citenamefont
  {Katanin}, \citenamefont {Antipov}, \citenamefont {Katsnelson}, \citenamefont
  {Lichtenstein}, \citenamefont {Rubtsov},\ and\ \citenamefont
  {Held}}]{Rohringer2018}%
  \BibitemOpen
  \bibfield  {author} {\bibinfo {author} {\bibfnamefont {G.}~\bibnamefont
  {Rohringer}}, \bibinfo {author} {\bibfnamefont {H.}~\bibnamefont
  {Hafermann}}, \bibinfo {author} {\bibfnamefont {A.}~\bibnamefont {Toschi}},
  \bibinfo {author} {\bibfnamefont {A.~A.}\ \bibnamefont {Katanin}}, \bibinfo
  {author} {\bibfnamefont {A.~E.}\ \bibnamefont {Antipov}}, \bibinfo {author}
  {\bibfnamefont {M.~I.}\ \bibnamefont {Katsnelson}}, \bibinfo {author}
  {\bibfnamefont {A.~I.}\ \bibnamefont {Lichtenstein}}, \bibinfo {author}
  {\bibfnamefont {A.~N.}\ \bibnamefont {Rubtsov}},\ and\ \bibinfo {author}
  {\bibfnamefont {K.}~\bibnamefont {Held}},\ }\href@noop {} {\bibfield
  {journal} {\bibinfo  {journal} {Rev. Mod. Phys.}\ }\textbf {\bibinfo {volume}
  {90}} (\bibinfo {year} {2018})}\BibitemShut {NoStop}%
\bibitem [{\citenamefont {Petocchi}\ \emph {et~al.}(2020)\citenamefont
  {Petocchi}, \citenamefont {Christiansson}, \citenamefont {Nilsson},
  \citenamefont {Aryasetiawan},\ and\ \citenamefont {Werner}}]{Petocchi2020}%
  \BibitemOpen
  \bibfield  {author} {\bibinfo {author} {\bibfnamefont {F.}~\bibnamefont
  {Petocchi}}, \bibinfo {author} {\bibfnamefont {V.}~\bibnamefont
  {Christiansson}}, \bibinfo {author} {\bibfnamefont {F.}~\bibnamefont
  {Nilsson}}, \bibinfo {author} {\bibfnamefont {F.}~\bibnamefont
  {Aryasetiawan}},\ and\ \bibinfo {author} {\bibfnamefont {P.}~\bibnamefont
  {Werner}},\ }\href@noop {} {\bibfield  {journal} {\bibinfo  {journal} {Phys.
  Rev. X}\ }\textbf {\bibinfo {volume} {10}} (\bibinfo {year}
  {2020})}\BibitemShut {NoStop}%
\bibitem [{\citenamefont {Knizia}\ and\ \citenamefont
  {Chan}(2012)}]{Knizia2012}%
  \BibitemOpen
  \bibfield  {author} {\bibinfo {author} {\bibfnamefont {G.}~\bibnamefont
  {Knizia}}\ and\ \bibinfo {author} {\bibfnamefont {G.~K.~L.}\ \bibnamefont
  {Chan}},\ }\href@noop {} {\bibfield  {journal} {\bibinfo  {journal} {Phys.
  Rev. Lett.}\ }\textbf {\bibinfo {volume} {109}},\ \bibinfo {pages} {1}
  (\bibinfo {year} {2012})}\BibitemShut {NoStop}%
\bibitem [{\citenamefont {Knizia}\ and\ \citenamefont
  {Chan}(2013)}]{Knizia2013}%
  \BibitemOpen
  \bibfield  {author} {\bibinfo {author} {\bibfnamefont {G.}~\bibnamefont
  {Knizia}}\ and\ \bibinfo {author} {\bibfnamefont {G.~K.~L.}\ \bibnamefont
  {Chan}},\ }\href@noop {} {\bibfield  {journal} {\bibinfo  {journal} {J. Chem.
  Theory Comput.}\ }\textbf {\bibinfo {volume} {9}},\ \bibinfo {pages} {1428}
  (\bibinfo {year} {2013})}\BibitemShut {NoStop}%
\bibitem [{\citenamefont {Chen}\ \emph {et~al.}(2014)\citenamefont {Chen},
  \citenamefont {Booth}, \citenamefont {Sharma}, \citenamefont {Knizia},\ and\
  \citenamefont {Chan}}]{Chen2014}%
  \BibitemOpen
  \bibfield  {author} {\bibinfo {author} {\bibfnamefont {Q.}~\bibnamefont
  {Chen}}, \bibinfo {author} {\bibfnamefont {G.~H.}\ \bibnamefont {Booth}},
  \bibinfo {author} {\bibfnamefont {S.}~\bibnamefont {Sharma}}, \bibinfo
  {author} {\bibfnamefont {G.}~\bibnamefont {Knizia}},\ and\ \bibinfo {author}
  {\bibfnamefont {G.~K.~L.}\ \bibnamefont {Chan}},\ }\href@noop {} {\bibfield
  {journal} {\bibinfo  {journal} {Phys. Rev. B - Condens. Matter Mater. Phys.}\
  }\textbf {\bibinfo {volume} {89}},\ \bibinfo {pages} {165134} (\bibinfo
  {year} {2014})}\BibitemShut {NoStop}%
\bibitem [{\citenamefont {Holmes}\ \emph {et~al.}(2016)\citenamefont {Holmes},
  \citenamefont {Tubman},\ and\ \citenamefont {Umrigar}}]{Holmes2016a}%
  \BibitemOpen
  \bibfield  {author} {\bibinfo {author} {\bibfnamefont {A.~A.}\ \bibnamefont
  {Holmes}}, \bibinfo {author} {\bibfnamefont {N.~M.}\ \bibnamefont {Tubman}},\
  and\ \bibinfo {author} {\bibfnamefont {C.~J.}\ \bibnamefont {Umrigar}},\
  }\href@noop {} {\bibfield  {journal} {\bibinfo  {journal} {J. Chem. Theory
  Comput.}\ }\textbf {\bibinfo {volume} {12}},\ \bibinfo {pages} {3674}
  (\bibinfo {year} {2016})}\BibitemShut {NoStop}%
\bibitem [{\citenamefont {Zheng}\ \emph
  {et~al.}(2017{\natexlab{a}})\citenamefont {Zheng}, \citenamefont {Kretchmer},
  \citenamefont {Shi}, \citenamefont {Zhang},\ and\ \citenamefont
  {Chan}}]{Zheng2017b}%
  \BibitemOpen
  \bibfield  {author} {\bibinfo {author} {\bibfnamefont {B.~X.}\ \bibnamefont
  {Zheng}}, \bibinfo {author} {\bibfnamefont {J.~S.}\ \bibnamefont
  {Kretchmer}}, \bibinfo {author} {\bibfnamefont {H.}~\bibnamefont {Shi}},
  \bibinfo {author} {\bibfnamefont {S.}~\bibnamefont {Zhang}},\ and\ \bibinfo
  {author} {\bibfnamefont {G.~K.~L.}\ \bibnamefont {Chan}},\ }\href@noop {}
  {\bibfield  {journal} {\bibinfo  {journal} {Phys. Rev. B}\ }\textbf {\bibinfo
  {volume} {95}},\ \bibinfo {pages} {045103} (\bibinfo {year}
  {2017}{\natexlab{a}})}\BibitemShut {NoStop}%
\bibitem [{\citenamefont {Pham}\ \emph {et~al.}(2018)\citenamefont {Pham},
  \citenamefont {Bernales},\ and\ \citenamefont {Gagliardi}}]{Pham2018}%
  \BibitemOpen
  \bibfield  {author} {\bibinfo {author} {\bibfnamefont {H.~Q.}\ \bibnamefont
  {Pham}}, \bibinfo {author} {\bibfnamefont {V.}~\bibnamefont {Bernales}},\
  and\ \bibinfo {author} {\bibfnamefont {L.}~\bibnamefont {Gagliardi}},\
  }\href@noop {} {\bibfield  {journal} {\bibinfo  {journal} {J. Chem. Theory
  Comput.}\ }\textbf {\bibinfo {volume} {14}},\ \bibinfo {pages} {1960}
  (\bibinfo {year} {2018})}\BibitemShut {NoStop}%
\bibitem [{\citenamefont {Cui}\ \emph {et~al.}(2020)\citenamefont {Cui},
  \citenamefont {Zhu},\ and\ \citenamefont {Chan}}]{Cui2020}%
  \BibitemOpen
  \bibfield  {author} {\bibinfo {author} {\bibfnamefont {Z.~H.}\ \bibnamefont
  {Cui}}, \bibinfo {author} {\bibfnamefont {T.}~\bibnamefont {Zhu}},\ and\
  \bibinfo {author} {\bibfnamefont {G.~K.~L.}\ \bibnamefont {Chan}},\
  }\href@noop {} {\bibfield  {journal} {\bibinfo  {journal} {J. Chem. Theory
  Comput.}\ }\textbf {\bibinfo {volume} {16}},\ \bibinfo {pages} {119}
  (\bibinfo {year} {2020})}\BibitemShut {NoStop}%
\bibitem [{\citenamefont {Fertitta}\ and\ \citenamefont
  {Booth}(2018)}]{Fertitta2018}%
  \BibitemOpen
  \bibfield  {author} {\bibinfo {author} {\bibfnamefont {E.}~\bibnamefont
  {Fertitta}}\ and\ \bibinfo {author} {\bibfnamefont {G.~H.}\ \bibnamefont
  {Booth}},\ }\href@noop {} {\bibfield  {journal} {\bibinfo  {journal} {Phys.
  Rev. B}\ }\textbf {\bibinfo {volume} {98}},\ \bibinfo {pages} {235132}
  (\bibinfo {year} {2018})}\BibitemShut {NoStop}%
\bibitem [{\citenamefont {Fertitta}\ and\ \citenamefont
  {Booth}(2019)}]{Fertitta2019}%
  \BibitemOpen
  \bibfield  {author} {\bibinfo {author} {\bibfnamefont {E.}~\bibnamefont
  {Fertitta}}\ and\ \bibinfo {author} {\bibfnamefont {G.~H.}\ \bibnamefont
  {Booth}},\ }\href@noop {} {\bibfield  {journal} {\bibinfo  {journal} {J.
  Chem. Phys.}\ }\textbf {\bibinfo {volume} {151}},\ \bibinfo {pages} {14115}
  (\bibinfo {year} {2019})}\BibitemShut {NoStop}%
\bibitem [{\citenamefont {Sriluckshmy}\ \emph {et~al.}(2021)\citenamefont
  {Sriluckshmy}, \citenamefont {Nusspickel}, \citenamefont {Fertitta},\ and\
  \citenamefont {Booth}}]{Sriluckshmy2021}%
  \BibitemOpen
  \bibfield  {author} {\bibinfo {author} {\bibfnamefont {P.~V.}\ \bibnamefont
  {Sriluckshmy}}, \bibinfo {author} {\bibfnamefont {M.}~\bibnamefont
  {Nusspickel}}, \bibinfo {author} {\bibfnamefont {E.}~\bibnamefont
  {Fertitta}},\ and\ \bibinfo {author} {\bibfnamefont {G.~H.}\ \bibnamefont
  {Booth}},\ }\href@noop {} {\bibfield  {journal} {\bibinfo  {journal} {Phys.
  Rev. B}\ }\textbf {\bibinfo {volume} {103}},\ \bibinfo {pages} {85131}
  (\bibinfo {year} {2021})}\BibitemShut {NoStop}%
\bibitem [{\citenamefont {LeBlanc}\ \emph {et~al.}(2015)\citenamefont
  {LeBlanc}, \citenamefont {Antipov}, \citenamefont {Becca}, \citenamefont
  {Bulik}, \citenamefont {Chan}, \citenamefont {Chung}, \citenamefont {Deng},
  \citenamefont {Ferrero}, \citenamefont {Henderson}, \citenamefont
  {Jim{\'{e}}nez-Hoyos}, \citenamefont {Kozik}, \citenamefont {Liu},
  \citenamefont {Millis}, \citenamefont {Prokof'ev}, \citenamefont {Qin},
  \citenamefont {Scuseria}, \citenamefont {Shi}, \citenamefont {Svistunov},
  \citenamefont {Tocchio}, \citenamefont {Tupitsyn}, \citenamefont {White},
  \citenamefont {Zhang}, \citenamefont {Zheng}, \citenamefont {Zhu},\ and\
  \citenamefont {Gull}}]{LeBlanc2015}%
  \BibitemOpen
  \bibfield  {author} {\bibinfo {author} {\bibfnamefont {J.~P.~F.}\
  \bibnamefont {LeBlanc}}, \bibinfo {author} {\bibfnamefont {A.~E.}\
  \bibnamefont {Antipov}}, \bibinfo {author} {\bibfnamefont {F.}~\bibnamefont
  {Becca}}, \bibinfo {author} {\bibfnamefont {I.~W.}\ \bibnamefont {Bulik}},
  \bibinfo {author} {\bibfnamefont {G.~K.-L.}\ \bibnamefont {Chan}}, \bibinfo
  {author} {\bibfnamefont {C.-M.}\ \bibnamefont {Chung}}, \bibinfo {author}
  {\bibfnamefont {Y.}~\bibnamefont {Deng}}, \bibinfo {author} {\bibfnamefont
  {M.}~\bibnamefont {Ferrero}}, \bibinfo {author} {\bibfnamefont {T.~M.}\
  \bibnamefont {Henderson}}, \bibinfo {author} {\bibfnamefont {C.~A.}\
  \bibnamefont {Jim{\'{e}}nez-Hoyos}}, \bibinfo {author} {\bibfnamefont
  {E.}~\bibnamefont {Kozik}}, \bibinfo {author} {\bibfnamefont {X.-W.}\
  \bibnamefont {Liu}}, \bibinfo {author} {\bibfnamefont {A.~J.}\ \bibnamefont
  {Millis}}, \bibinfo {author} {\bibfnamefont {N.~V.}\ \bibnamefont
  {Prokof'ev}}, \bibinfo {author} {\bibfnamefont {M.}~\bibnamefont {Qin}},
  \bibinfo {author} {\bibfnamefont {G.~E.}\ \bibnamefont {Scuseria}}, \bibinfo
  {author} {\bibfnamefont {H.}~\bibnamefont {Shi}}, \bibinfo {author}
  {\bibfnamefont {B.~V.}\ \bibnamefont {Svistunov}}, \bibinfo {author}
  {\bibfnamefont {L.~F.}\ \bibnamefont {Tocchio}}, \bibinfo {author}
  {\bibfnamefont {I.~S.}\ \bibnamefont {Tupitsyn}}, \bibinfo {author}
  {\bibfnamefont {S.~R.}\ \bibnamefont {White}}, \bibinfo {author}
  {\bibfnamefont {S.}~\bibnamefont {Zhang}}, \bibinfo {author} {\bibfnamefont
  {B.-X.}\ \bibnamefont {Zheng}}, \bibinfo {author} {\bibfnamefont
  {Z.}~\bibnamefont {Zhu}},\ and\ \bibinfo {author} {\bibfnamefont
  {E.}~\bibnamefont {Gull}},\ }\href@noop {} {\bibfield  {journal} {\bibinfo
  {journal} {Phys. Rev. X}\ }\textbf {\bibinfo {volume} {5}} (\bibinfo {year}
  {2015})}\BibitemShut {NoStop}%
\bibitem [{\citenamefont {Zheng}\ \emph
  {et~al.}(2017{\natexlab{b}})\citenamefont {Zheng}, \citenamefont {Chung},
  \citenamefont {Corboz}, \citenamefont {Ehlers}, \citenamefont {Qin},
  \citenamefont {Noack}, \citenamefont {Shi}, \citenamefont {White},
  \citenamefont {Zhang},\ and\ \citenamefont {Chan}}]{Zheng2017a}%
  \BibitemOpen
  \bibfield  {author} {\bibinfo {author} {\bibfnamefont {B.~X.}\ \bibnamefont
  {Zheng}}, \bibinfo {author} {\bibfnamefont {C.~M.}\ \bibnamefont {Chung}},
  \bibinfo {author} {\bibfnamefont {P.}~\bibnamefont {Corboz}}, \bibinfo
  {author} {\bibfnamefont {G.}~\bibnamefont {Ehlers}}, \bibinfo {author}
  {\bibfnamefont {M.~P.}\ \bibnamefont {Qin}}, \bibinfo {author} {\bibfnamefont
  {R.~M.}\ \bibnamefont {Noack}}, \bibinfo {author} {\bibfnamefont
  {H.}~\bibnamefont {Shi}}, \bibinfo {author} {\bibfnamefont {S.~R.}\
  \bibnamefont {White}}, \bibinfo {author} {\bibfnamefont {S.}~\bibnamefont
  {Zhang}},\ and\ \bibinfo {author} {\bibfnamefont {G.~K.~L.}\ \bibnamefont
  {Chan}},\ }\href@noop {} {\bibfield  {journal} {\bibinfo  {journal} {Science
  (80-. ).}\ }\textbf {\bibinfo {volume} {358}},\ \bibinfo {pages} {1155}
  (\bibinfo {year} {2017}{\natexlab{b}})}\BibitemShut {NoStop}%
\bibitem [{\citenamefont {Gell-Mann}\ and\ \citenamefont
  {Brueckner}(1957)}]{Gell-Mann1957}%
  \BibitemOpen
  \bibfield  {author} {\bibinfo {author} {\bibfnamefont {M.}~\bibnamefont
  {Gell-Mann}}\ and\ \bibinfo {author} {\bibfnamefont {K.~A.}\ \bibnamefont
  {Brueckner}},\ }\href@noop {} {\bibfield  {journal} {\bibinfo  {journal}
  {Phys. Rev.}\ }\textbf {\bibinfo {volume} {106}},\ \bibinfo {pages} {364}
  (\bibinfo {year} {1957})}\BibitemShut {NoStop}%
\bibitem [{\citenamefont {Eguiluz}(1983)}]{Eguiluz1983}%
  \BibitemOpen
  \bibfield  {author} {\bibinfo {author} {\bibfnamefont {A.~G.}\ \bibnamefont
  {Eguiluz}},\ }\href@noop {} {\bibfield  {journal} {\bibinfo  {journal} {Phys.
  Rev. Lett.}\ }\textbf {\bibinfo {volume} {51}},\ \bibinfo {pages} {1907}
  (\bibinfo {year} {1983})}\BibitemShut {NoStop}%
\bibitem [{\citenamefont {Li}\ \emph {et~al.}(1992)\citenamefont {Li},
  \citenamefont {{Das Sarma}},\ and\ \citenamefont {Joynt}}]{Li1992}%
  \BibitemOpen
  \bibfield  {author} {\bibinfo {author} {\bibfnamefont {Q.~P.}\ \bibnamefont
  {Li}}, \bibinfo {author} {\bibfnamefont {S.}~\bibnamefont {{Das Sarma}}},\
  and\ \bibinfo {author} {\bibfnamefont {R.}~\bibnamefont {Joynt}},\
  }\href@noop {} {\bibfield  {journal} {\bibinfo  {journal} {Phys. Rev. B}\
  }\textbf {\bibinfo {volume} {45}},\ \bibinfo {pages} {13713} (\bibinfo {year}
  {1992})}\BibitemShut {NoStop}%
\bibitem [{\citenamefont {{Garc{\'{i}}a De Abajo}}\ and\ \citenamefont
  {Echenique}(1993)}]{GarciaDeAbajo1993}%
  \BibitemOpen
  \bibfield  {author} {\bibinfo {author} {\bibfnamefont {F.~J.}\ \bibnamefont
  {{Garc{\'{i}}a De Abajo}}}\ and\ \bibinfo {author} {\bibfnamefont {P.~M.}\
  \bibnamefont {Echenique}},\ }\href@noop {} {\bibfield  {journal} {\bibinfo
  {journal} {Phys. Rev. B}\ }\textbf {\bibinfo {volume} {48}},\ \bibinfo
  {pages} {13399} (\bibinfo {year} {1993})}\BibitemShut {NoStop}%
\bibitem [{\citenamefont {Polini}\ \emph {et~al.}(2008)\citenamefont {Polini},
  \citenamefont {Asgari}, \citenamefont {Borghi}, \citenamefont {Barlas},
  \citenamefont {Pereg-Barnea},\ and\ \citenamefont {MacDonald}}]{Polini2008}%
  \BibitemOpen
  \bibfield  {author} {\bibinfo {author} {\bibfnamefont {M.}~\bibnamefont
  {Polini}}, \bibinfo {author} {\bibfnamefont {R.}~\bibnamefont {Asgari}},
  \bibinfo {author} {\bibfnamefont {G.}~\bibnamefont {Borghi}}, \bibinfo
  {author} {\bibfnamefont {Y.}~\bibnamefont {Barlas}}, \bibinfo {author}
  {\bibfnamefont {T.}~\bibnamefont {Pereg-Barnea}},\ and\ \bibinfo {author}
  {\bibfnamefont {A.~H.}\ \bibnamefont {MacDonald}},\ }\href@noop {} {\bibfield
   {journal} {\bibinfo  {journal} {Phys. Rev. B - Condens. Matter Mater.
  Phys.}\ }\textbf {\bibinfo {volume} {77}},\ \bibinfo {pages} {081411}
  (\bibinfo {year} {2008})}\BibitemShut {NoStop}%
\bibitem [{\citenamefont {Ichikawa}(2016)}]{Ichikawa2016}%
  \BibitemOpen
  \bibfield  {author} {\bibinfo {author} {\bibfnamefont {M.}~\bibnamefont
  {Ichikawa}},\ }\href@noop {} {\bibfield  {journal} {\bibinfo  {journal}
  {Condens. Matter}\ }\textbf {\bibinfo {volume} {1}},\ \bibinfo {pages} {1}
  (\bibinfo {year} {2016})}\BibitemShut {NoStop}%
\bibitem [{\citenamefont {Chen}\ \emph {et~al.}(2017)\citenamefont {Chen},
  \citenamefont {Voora}, \citenamefont {Agee}, \citenamefont {Balasubramani},\
  and\ \citenamefont {Furche}}]{Chen2017}%
  \BibitemOpen
  \bibfield  {author} {\bibinfo {author} {\bibfnamefont {G.~P.}\ \bibnamefont
  {Chen}}, \bibinfo {author} {\bibfnamefont {V.~K.}\ \bibnamefont {Voora}},
  \bibinfo {author} {\bibfnamefont {M.~M.}\ \bibnamefont {Agee}}, \bibinfo
  {author} {\bibfnamefont {G.}~\bibnamefont {Balasubramani}},\ and\ \bibinfo
  {author} {\bibfnamefont {F.}~\bibnamefont {Furche}},\ }\href@noop {}
  {\bibfield  {journal} {\bibinfo  {journal} {Annu. Rev. Phys. Chem.}\ }\textbf
  {\bibinfo {volume} {68}},\ \bibinfo {pages} {421} (\bibinfo {year}
  {2017})}\BibitemShut {NoStop}%
\bibitem [{\citenamefont {Ren}\ \emph {et~al.}(2012)\citenamefont {Ren},
  \citenamefont {Rinke}, \citenamefont {Joas},\ and\ \citenamefont
  {Scheffler}}]{Ren2012}%
  \BibitemOpen
  \bibfield  {author} {\bibinfo {author} {\bibfnamefont {X.}~\bibnamefont
  {Ren}}, \bibinfo {author} {\bibfnamefont {P.}~\bibnamefont {Rinke}}, \bibinfo
  {author} {\bibfnamefont {C.}~\bibnamefont {Joas}},\ and\ \bibinfo {author}
  {\bibfnamefont {M.}~\bibnamefont {Scheffler}},\ }\href@noop {} {\bibfield
  {journal} {\bibinfo  {journal} {J. Mater. Sci.}\ }\textbf {\bibinfo {volume}
  {47}},\ \bibinfo {pages} {7447} (\bibinfo {year} {2012})}\BibitemShut
  {NoStop}%
\bibitem [{\citenamefont {Hedin}(1965)}]{Hedin1965}%
  \BibitemOpen
  \bibfield  {author} {\bibinfo {author} {\bibfnamefont {L.}~\bibnamefont
  {Hedin}},\ }\href@noop {} {\bibfield  {journal} {\bibinfo  {journal} {Phys.
  Rev.}\ }\textbf {\bibinfo {volume} {139}},\ \bibinfo {pages} {A796} (\bibinfo
  {year} {1965})}\BibitemShut {NoStop}%
\bibitem [{\citenamefont {Springer}\ and\ \citenamefont
  {Aryasetiawan}(1998)}]{Springer1998}%
  \BibitemOpen
  \bibfield  {author} {\bibinfo {author} {\bibfnamefont {M.}~\bibnamefont
  {Springer}}\ and\ \bibinfo {author} {\bibfnamefont {F.}~\bibnamefont
  {Aryasetiawan}},\ }\href@noop {} {\bibfield  {journal} {\bibinfo  {journal}
  {Phys. Rev. B Condens. Matter}\ }\textbf {\bibinfo {volume} {57}},\ \bibinfo
  {pages} {4364} (\bibinfo {year} {1998})}\BibitemShut {NoStop}%
\bibitem [{Miy(2008)}]{Miyake2008}%
  \BibitemOpen
  \href@noop {} {\bibfield  {journal} {\bibinfo  {journal} {Phys. Rev. B
  Condens. Matter}\ }\textbf {\bibinfo {volume} {77}} (\bibinfo {year}
  {2008})}\BibitemShut {NoStop}%
\bibitem [{\citenamefont {Miyake}\ \emph {et~al.}(2009)\citenamefont {Miyake},
  \citenamefont {Aryasetiawan},\ and\ \citenamefont {Imada}}]{Miyake2009}%
  \BibitemOpen
  \bibfield  {author} {\bibinfo {author} {\bibfnamefont {T.}~\bibnamefont
  {Miyake}}, \bibinfo {author} {\bibfnamefont {F.}~\bibnamefont
  {Aryasetiawan}},\ and\ \bibinfo {author} {\bibfnamefont {M.}~\bibnamefont
  {Imada}},\ }\href@noop {} {\bibfield  {journal} {\bibinfo  {journal} {Phys.
  Rev. B Condens. Matter}\ }\textbf {\bibinfo {volume} {80}} (\bibinfo {year}
  {2009})}\BibitemShut {NoStop}%
\bibitem [{\citenamefont {Jiang}\ \emph {et~al.}(2010)\citenamefont {Jiang},
  \citenamefont {Gomez-Abal}, \citenamefont {Rinke},\ and\ \citenamefont
  {Scheffler}}]{Jiang2010}%
  \BibitemOpen
  \bibfield  {author} {\bibinfo {author} {\bibfnamefont {H.}~\bibnamefont
  {Jiang}}, \bibinfo {author} {\bibfnamefont {R.~I.}\ \bibnamefont
  {Gomez-Abal}}, \bibinfo {author} {\bibfnamefont {P.}~\bibnamefont {Rinke}},\
  and\ \bibinfo {author} {\bibfnamefont {M.}~\bibnamefont {Scheffler}},\
  }\href@noop {} {\bibfield  {journal} {\bibinfo  {journal} {Phys. Rev. B
  Condens. Matter}\ }\textbf {\bibinfo {volume} {82}} (\bibinfo {year}
  {2010})}\BibitemShut {NoStop}%
\bibitem [{\citenamefont {{\c{S}}a{\c{s}}{\i}o{\u{g}}lu}\ \emph
  {et~al.}(2011)\citenamefont {{\c{S}}a{\c{s}}{\i}o{\u{g}}lu}, \citenamefont
  {Friedrich},\ and\ \citenamefont {Bl{\"u}gel}}]{Sasioglu2011}%
  \BibitemOpen
  \bibfield  {author} {\bibinfo {author} {\bibfnamefont {E.}~\bibnamefont
  {{\c{S}}a{\c{s}}{\i}o{\u{g}}lu}}, \bibinfo {author} {\bibfnamefont
  {C.}~\bibnamefont {Friedrich}},\ and\ \bibinfo {author} {\bibfnamefont
  {S.}~\bibnamefont {Bl{\"u}gel}},\ }\href@noop {} {\bibfield  {journal}
  {\bibinfo  {journal} {Phys. Rev. B}\ }\textbf {\bibinfo {volume} {83}},\
  \bibinfo {pages} {121101} (\bibinfo {year} {2011})}\BibitemShut {NoStop}%
\bibitem [{\citenamefont {Casida}(1995)}]{Casida1995}%
  \BibitemOpen
  \bibfield  {author} {\bibinfo {author} {\bibfnamefont {M.~E.}\ \bibnamefont
  {Casida}},\ }in\ \href@noop {} {\emph {\bibinfo {booktitle} {Recent Advances
  In Density Functional Methods: (Part I)}}}\ (\bibinfo  {publisher} {World
  Scientific},\ \bibinfo {year} {1995})\ pp.\ \bibinfo {pages}
  {155--192}\BibitemShut {NoStop}%
\bibitem [{\citenamefont {Casida}\ \emph {et~al.}(1998)\citenamefont {Casida},
  \citenamefont {Jamorski}, \citenamefont {Casida},\ and\ \citenamefont
  {Salahub}}]{Casida1998}%
  \BibitemOpen
  \bibfield  {author} {\bibinfo {author} {\bibfnamefont {M.~E.}\ \bibnamefont
  {Casida}}, \bibinfo {author} {\bibfnamefont {C.}~\bibnamefont {Jamorski}},
  \bibinfo {author} {\bibfnamefont {K.~C.}\ \bibnamefont {Casida}},\ and\
  \bibinfo {author} {\bibfnamefont {D.~R.}\ \bibnamefont {Salahub}},\
  }\href@noop {} {\bibfield  {journal} {\bibinfo  {journal} {J. Chem. Phys.}\
  }\textbf {\bibinfo {volume} {108}},\ \bibinfo {pages} {4439} (\bibinfo {year}
  {1998})}\BibitemShut {NoStop}%
\bibitem [{\citenamefont {Furche}(2001)}]{Furche2001a}%
  \BibitemOpen
  \bibfield  {author} {\bibinfo {author} {\bibfnamefont {F.}~\bibnamefont
  {Furche}},\ }\href@noop {} {\bibfield  {journal} {\bibinfo  {journal} {J.
  Chem. Phys.}\ }\textbf {\bibinfo {volume} {114}},\ \bibinfo {pages} {5982}
  (\bibinfo {year} {2001})}\BibitemShut {NoStop}%
\bibitem [{\citenamefont {{\'{A}}ngy{\'{a}}n}\ \emph
  {et~al.}(2011)\citenamefont {{\'{A}}ngy{\'{a}}n}, \citenamefont {Liu},
  \citenamefont {Toulouse},\ and\ \citenamefont {Jansen}}]{Angyan2011}%
  \BibitemOpen
  \bibfield  {author} {\bibinfo {author} {\bibfnamefont {J.~G.}\ \bibnamefont
  {{\'{A}}ngy{\'{a}}n}}, \bibinfo {author} {\bibfnamefont {R.~F.}\ \bibnamefont
  {Liu}}, \bibinfo {author} {\bibfnamefont {J.}~\bibnamefont {Toulouse}},\ and\
  \bibinfo {author} {\bibfnamefont {G.}~\bibnamefont {Jansen}},\ }\href
  {https://doi.org/10.1021/ct200501r} {\bibfield  {journal} {\bibinfo
  {journal} {J. Chem. Theory Comput.}\ }\textbf {\bibinfo {volume} {7}},\
  \bibinfo {pages} {3116} (\bibinfo {year} {2011})},\ \Eprint
  {https://arxiv.org/abs/1404.1663} {arXiv:1404.1663} \BibitemShut {NoStop}%
\bibitem [{\citenamefont {Wouters}\ \emph {et~al.}(2016)\citenamefont
  {Wouters}, \citenamefont {Jim{\'{e}}nez-Hoyos}, \citenamefont {Sun},\ and\
  \citenamefont {Chan}}]{Wouters2016}%
  \BibitemOpen
  \bibfield  {author} {\bibinfo {author} {\bibfnamefont {S.}~\bibnamefont
  {Wouters}}, \bibinfo {author} {\bibfnamefont {C.~A.}\ \bibnamefont
  {Jim{\'{e}}nez-Hoyos}}, \bibinfo {author} {\bibfnamefont {Q.}~\bibnamefont
  {Sun}},\ and\ \bibinfo {author} {\bibfnamefont {G.~K.}\ \bibnamefont
  {Chan}},\ }\href@noop {} {\bibfield  {journal} {\bibinfo  {journal} {J. Chem.
  Theory Comput.}\ }\textbf {\bibinfo {volume} {12}},\ \bibinfo {pages} {2706}
  (\bibinfo {year} {2016})}\BibitemShut {NoStop}%
\bibitem [{\citenamefont {Botti}\ \emph {et~al.}(2007)\citenamefont {Botti},
  \citenamefont {Schindlmayr}, \citenamefont {{Del Sole}},\ and\ \citenamefont
  {Reining}}]{Botti2007}%
  \BibitemOpen
  \bibfield  {author} {\bibinfo {author} {\bibfnamefont {S.}~\bibnamefont
  {Botti}}, \bibinfo {author} {\bibfnamefont {A.}~\bibnamefont {Schindlmayr}},
  \bibinfo {author} {\bibfnamefont {R.}~\bibnamefont {{Del Sole}}},\ and\
  \bibinfo {author} {\bibfnamefont {L.}~\bibnamefont {Reining}},\ }\href@noop
  {} {\bibfield  {journal} {\bibinfo  {journal} {Reports Prog. Phys.}\ }\textbf
  {\bibinfo {volume} {70}},\ \bibinfo {pages} {357} (\bibinfo {year}
  {2007})}\BibitemShut {NoStop}%
\bibitem [{\citenamefont {Wu}\ \emph {et~al.}(2019)\citenamefont {Wu},
  \citenamefont {Cui}, \citenamefont {Tong}, \citenamefont {Lindsey},
  \citenamefont {Chan},\ and\ \citenamefont {Lin}}]{Wu2019}%
  \BibitemOpen
  \bibfield  {author} {\bibinfo {author} {\bibfnamefont {X.}~\bibnamefont
  {Wu}}, \bibinfo {author} {\bibfnamefont {Z.~H.}\ \bibnamefont {Cui}},
  \bibinfo {author} {\bibfnamefont {Y.}~\bibnamefont {Tong}}, \bibinfo {author}
  {\bibfnamefont {M.}~\bibnamefont {Lindsey}}, \bibinfo {author} {\bibfnamefont
  {G.~K.~L.}\ \bibnamefont {Chan}},\ and\ \bibinfo {author} {\bibfnamefont
  {L.}~\bibnamefont {Lin}},\ }\href@noop {} {\bibfield  {journal} {\bibinfo
  {journal} {J. Chem. Phys.}\ }\textbf {\bibinfo {volume} {151}} (\bibinfo
  {year} {2019})}\BibitemShut {NoStop}%
\bibitem [{\citenamefont {Wu}\ \emph {et~al.}(2020)\citenamefont {Wu},
  \citenamefont {Lindsey}, \citenamefont {Zhou}, \citenamefont {Tong},\ and\
  \citenamefont {Lin}}]{Wu2020}%
  \BibitemOpen
  \bibfield  {author} {\bibinfo {author} {\bibfnamefont {X.}~\bibnamefont
  {Wu}}, \bibinfo {author} {\bibfnamefont {M.}~\bibnamefont {Lindsey}},
  \bibinfo {author} {\bibfnamefont {T.}~\bibnamefont {Zhou}}, \bibinfo {author}
  {\bibfnamefont {Y.}~\bibnamefont {Tong}},\ and\ \bibinfo {author}
  {\bibfnamefont {L.}~\bibnamefont {Lin}},\ }\href@noop {} {\bibfield
  {journal} {\bibinfo  {journal} {Phys. Rev. B}\ }\textbf {\bibinfo {volume}
  {102}} (\bibinfo {year} {2020})}\BibitemShut {NoStop}%
\bibitem [{\citenamefont {O'Donoghue}\ \emph {et~al.}(2016)\citenamefont
  {O'Donoghue}, \citenamefont {Chu}, \citenamefont {Parikh},\ and\
  \citenamefont {Boyd}}]{ocpb:16}%
  \BibitemOpen
  \bibfield  {author} {\bibinfo {author} {\bibfnamefont {B.}~\bibnamefont
  {O'Donoghue}}, \bibinfo {author} {\bibfnamefont {E.}~\bibnamefont {Chu}},
  \bibinfo {author} {\bibfnamefont {N.}~\bibnamefont {Parikh}},\ and\ \bibinfo
  {author} {\bibfnamefont {S.}~\bibnamefont {Boyd}},\ }\href
  {http://stanford.edu/~boyd/papers/scs.html} {\bibfield  {journal} {\bibinfo
  {journal} {Journal of Optimization Theory and Applications}\ }\textbf
  {\bibinfo {volume} {169}},\ \bibinfo {pages} {1042} (\bibinfo {year}
  {2016})}\BibitemShut {NoStop}%
\bibitem [{\citenamefont {O'Donoghue}\ \emph {et~al.}(2019)\citenamefont
  {O'Donoghue}, \citenamefont {Chu}, \citenamefont {Parikh},\ and\
  \citenamefont {Boyd}}]{scs}%
  \BibitemOpen
  \bibfield  {author} {\bibinfo {author} {\bibfnamefont {B.}~\bibnamefont
  {O'Donoghue}}, \bibinfo {author} {\bibfnamefont {E.}~\bibnamefont {Chu}},
  \bibinfo {author} {\bibfnamefont {N.}~\bibnamefont {Parikh}},\ and\ \bibinfo
  {author} {\bibfnamefont {S.}~\bibnamefont {Boyd}},\ }\href@noop {} {\bibinfo
  {title} {{SCS}: Splitting conic solver, version 2.1.4}},\ \bibinfo
  {howpublished} {\url{https://github.com/cvxgrp/scs}} (\bibinfo {year}
  {2019})\BibitemShut {NoStop}%
\bibitem [{\citenamefont {Diamond}\ and\ \citenamefont
  {Boyd}(2016)}]{diamond2016cvxpy}%
  \BibitemOpen
  \bibfield  {author} {\bibinfo {author} {\bibfnamefont {S.}~\bibnamefont
  {Diamond}}\ and\ \bibinfo {author} {\bibfnamefont {S.}~\bibnamefont {Boyd}},\
  }\href@noop {} {\bibfield  {journal} {\bibinfo  {journal} {Journal of Machine
  Learning Research}\ }\textbf {\bibinfo {volume} {17}},\ \bibinfo {pages} {1}
  (\bibinfo {year} {2016})}\BibitemShut {NoStop}%
\bibitem [{\citenamefont {Agrawal}\ \emph {et~al.}(2018)\citenamefont
  {Agrawal}, \citenamefont {Verschueren}, \citenamefont {Diamond},\ and\
  \citenamefont {Boyd}}]{agrawal2018rewriting}%
  \BibitemOpen
  \bibfield  {author} {\bibinfo {author} {\bibfnamefont {A.}~\bibnamefont
  {Agrawal}}, \bibinfo {author} {\bibfnamefont {R.}~\bibnamefont
  {Verschueren}}, \bibinfo {author} {\bibfnamefont {S.}~\bibnamefont
  {Diamond}},\ and\ \bibinfo {author} {\bibfnamefont {S.}~\bibnamefont
  {Boyd}},\ }\href@noop {} {\bibfield  {journal} {\bibinfo  {journal} {Journal
  of Control and Decision}\ }\textbf {\bibinfo {volume} {5}},\ \bibinfo {pages}
  {42} (\bibinfo {year} {2018})}\BibitemShut {NoStop}%
\bibitem [{\citenamefont {Sengupta}\ \emph {et~al.}(2002)\citenamefont
  {Sengupta}, \citenamefont {Sandvik},\ and\ \citenamefont
  {Campbell}}]{Sengupta2002}%
  \BibitemOpen
  \bibfield  {author} {\bibinfo {author} {\bibfnamefont {P.}~\bibnamefont
  {Sengupta}}, \bibinfo {author} {\bibfnamefont {A.~W.}\ \bibnamefont
  {Sandvik}},\ and\ \bibinfo {author} {\bibfnamefont {D.~K.}\ \bibnamefont
  {Campbell}},\ }\href@noop {} {\bibfield  {journal} {\bibinfo  {journal}
  {Phys. Rev. B Condens. Matter}\ }\textbf {\bibinfo {volume} {65}},\ \bibinfo
  {pages} {1551131} (\bibinfo {year} {2002})}\BibitemShut {NoStop}%
\bibitem [{\citenamefont {Tsuchiizu}\ and\ \citenamefont
  {Furusaki}(2002)}]{Tsuchiizu2002}%
  \BibitemOpen
  \bibfield  {author} {\bibinfo {author} {\bibfnamefont {M.}~\bibnamefont
  {Tsuchiizu}}\ and\ \bibinfo {author} {\bibfnamefont {A.}~\bibnamefont
  {Furusaki}},\ }\href@noop {} {\bibfield  {journal} {\bibinfo  {journal}
  {Phys. Rev. Lett.}\ }\textbf {\bibinfo {volume} {88}},\ \bibinfo {pages} {4}
  (\bibinfo {year} {2002})}\BibitemShut {NoStop}%
\bibitem [{\citenamefont {Jeckelmann}(2002)}]{Jeckelmann2002}%
  \BibitemOpen
  \bibfield  {author} {\bibinfo {author} {\bibfnamefont {E.}~\bibnamefont
  {Jeckelmann}},\ }\href@noop {} {\bibfield  {journal} {\bibinfo  {journal}
  {Phys. Rev. Lett.}\ }\textbf {\bibinfo {volume} {89}} (\bibinfo {year}
  {2002})}\BibitemShut {NoStop}%
\bibitem [{\citenamefont {Ejima}\ and\ \citenamefont
  {Nishimoto}(2007)}]{Ejima2007}%
  \BibitemOpen
  \bibfield  {author} {\bibinfo {author} {\bibfnamefont {S.}~\bibnamefont
  {Ejima}}\ and\ \bibinfo {author} {\bibfnamefont {S.}~\bibnamefont
  {Nishimoto}},\ }\href@noop {} {\bibfield  {journal} {\bibinfo  {journal}
  {Phys. Rev. Lett.}\ }\textbf {\bibinfo {volume} {99}} (\bibinfo {year}
  {2007})}\BibitemShut {NoStop}%
\bibitem [{\citenamefont {Lanat{\`{a}}}(2020)}]{Lanata2020}%
  \BibitemOpen
  \bibfield  {author} {\bibinfo {author} {\bibfnamefont {N.}~\bibnamefont
  {Lanat{\`{a}}}},\ }\href@noop {} {\bibfield  {journal} {\bibinfo  {journal}
  {Phys. Rev. B}\ }\textbf {\bibinfo {volume} {102}},\ \bibinfo {pages}
  {115115} (\bibinfo {year} {2020})}\BibitemShut {NoStop}%
\bibitem [{\citenamefont {Lee}\ \emph {et~al.}(2019)\citenamefont {Lee},
  \citenamefont {Ayral}, \citenamefont {Yao}, \citenamefont {Lanata},\ and\
  \citenamefont {Kotliar}}]{Lee2019}%
  \BibitemOpen
  \bibfield  {author} {\bibinfo {author} {\bibfnamefont {T.~H.}\ \bibnamefont
  {Lee}}, \bibinfo {author} {\bibfnamefont {T.}~\bibnamefont {Ayral}}, \bibinfo
  {author} {\bibfnamefont {Y.~X.}\ \bibnamefont {Yao}}, \bibinfo {author}
  {\bibfnamefont {N.}~\bibnamefont {Lanata}},\ and\ \bibinfo {author}
  {\bibfnamefont {G.}~\bibnamefont {Kotliar}},\ }\href@noop {} {\bibfield
  {journal} {\bibinfo  {journal} {Phys. Rev. B}\ }\textbf {\bibinfo {volume}
  {99}},\ \bibinfo {pages} {115129} (\bibinfo {year} {2019})}\BibitemShut
  {NoStop}%
\bibitem [{\citenamefont {Nakamura}(1999)}]{Nakamura1999}%
  \BibitemOpen
  \bibfield  {author} {\bibinfo {author} {\bibfnamefont {M.}~\bibnamefont
  {Nakamura}},\ }\href@noop {} {\bibfield  {journal} {\bibinfo  {journal}
  {Journal of the Physical Society of Japan}\ }\textbf {\bibinfo {volume}
  {68}},\ \bibinfo {pages} {3123} (\bibinfo {year} {1999})}\BibitemShut
  {NoStop}%
\bibitem [{\citenamefont {Liu}\ and\ \citenamefont {Wang}(2011)}]{Liu2011}%
  \BibitemOpen
  \bibfield  {author} {\bibinfo {author} {\bibfnamefont {G.~H.}\ \bibnamefont
  {Liu}}\ and\ \bibinfo {author} {\bibfnamefont {C.~H.}\ \bibnamefont {Wang}},\
  }\href@noop {} {\bibfield  {journal} {\bibinfo  {journal} {Commun. Theor.
  Phys.}\ }\textbf {\bibinfo {volume} {55}},\ \bibinfo {pages} {702} (\bibinfo
  {year} {2011})}\BibitemShut {NoStop}%
\bibitem [{\citenamefont {Kim}\ \emph {et~al.}(2004)\citenamefont {Kim},
  \citenamefont {Hill}, \citenamefont {Benthien}, \citenamefont {Essler},
  \citenamefont {Jeckelmann}, \citenamefont {Choi}, \citenamefont {Non},
  \citenamefont {Motoyama}, \citenamefont {Kojima}, \citenamefont {Uchida},
  \citenamefont {Casa},\ and\ \citenamefont {Gog}}]{Kim2004}%
  \BibitemOpen
  \bibfield  {author} {\bibinfo {author} {\bibfnamefont {Y.~J.}\ \bibnamefont
  {Kim}}, \bibinfo {author} {\bibfnamefont {J.~P.}\ \bibnamefont {Hill}},
  \bibinfo {author} {\bibfnamefont {H.}~\bibnamefont {Benthien}}, \bibinfo
  {author} {\bibfnamefont {F.~H.}\ \bibnamefont {Essler}}, \bibinfo {author}
  {\bibfnamefont {E.}~\bibnamefont {Jeckelmann}}, \bibinfo {author}
  {\bibfnamefont {H.~S.}\ \bibnamefont {Choi}}, \bibinfo {author}
  {\bibfnamefont {T.~W.}\ \bibnamefont {Non}}, \bibinfo {author} {\bibfnamefont
  {N.}~\bibnamefont {Motoyama}}, \bibinfo {author} {\bibfnamefont {K.~M.}\
  \bibnamefont {Kojima}}, \bibinfo {author} {\bibfnamefont {S.}~\bibnamefont
  {Uchida}}, \bibinfo {author} {\bibfnamefont {D.}~\bibnamefont {Casa}},\ and\
  \bibinfo {author} {\bibfnamefont {T.}~\bibnamefont {Gog}},\ }\href@noop {}
  {\bibfield  {journal} {\bibinfo  {journal} {Phys. Rev. Lett.}\ }\textbf
  {\bibinfo {volume} {92}} (\bibinfo {year} {2004})}\BibitemShut {NoStop}%
\bibitem [{\citenamefont {Benthien}\ and\ \citenamefont
  {Jeckelmann}(2007)}]{Benthien2007}%
  \BibitemOpen
  \bibfield  {author} {\bibinfo {author} {\bibfnamefont {H.}~\bibnamefont
  {Benthien}}\ and\ \bibinfo {author} {\bibfnamefont {E.}~\bibnamefont
  {Jeckelmann}},\ }\href@noop {} {\bibfield  {journal} {\bibinfo  {journal}
  {Phys. Rev. B Condens. Matter}\ }\textbf {\bibinfo {volume} {75}} (\bibinfo
  {year} {2007})}\BibitemShut {NoStop}%
\bibitem [{\citenamefont {Onari}\ \emph {et~al.}(2004)\citenamefont {Onari},
  \citenamefont {Arita}, \citenamefont {Kuroki},\ and\ \citenamefont
  {Aoki}}]{Onari2004}%
  \BibitemOpen
  \bibfield  {author} {\bibinfo {author} {\bibfnamefont {S.}~\bibnamefont
  {Onari}}, \bibinfo {author} {\bibfnamefont {R.}~\bibnamefont {Arita}},
  \bibinfo {author} {\bibfnamefont {K.}~\bibnamefont {Kuroki}},\ and\ \bibinfo
  {author} {\bibfnamefont {H.}~\bibnamefont {Aoki}},\ }\href@noop {} {\bibfield
   {journal} {\bibinfo  {journal} {Phys. Rev. B - Condens. Matter Mater.
  Phys.}\ }\textbf {\bibinfo {volume} {70}},\ \bibinfo {pages} {094523}
  (\bibinfo {year} {2004})}\BibitemShut {NoStop}%
\bibitem [{\citenamefont {{Van Loon}}\ \emph {et~al.}(2014)\citenamefont {{Van
  Loon}}, \citenamefont {Lichtenstein}, \citenamefont {Katsnelson},
  \citenamefont {Parcollet},\ and\ \citenamefont {Hafermann}}]{VanLoon2014}%
  \BibitemOpen
  \bibfield  {author} {\bibinfo {author} {\bibfnamefont {E.~G.}\ \bibnamefont
  {{Van Loon}}}, \bibinfo {author} {\bibfnamefont {A.~I.}\ \bibnamefont
  {Lichtenstein}}, \bibinfo {author} {\bibfnamefont {M.~I.}\ \bibnamefont
  {Katsnelson}}, \bibinfo {author} {\bibfnamefont {O.}~\bibnamefont
  {Parcollet}},\ and\ \bibinfo {author} {\bibfnamefont {H.}~\bibnamefont
  {Hafermann}},\ }\href@noop {} {\bibfield  {journal} {\bibinfo  {journal}
  {Phys. Rev. B - Condens. Matter Mater. Phys.}\ }\textbf {\bibinfo {volume}
  {90}},\ \bibinfo {pages} {235135} (\bibinfo {year} {2014})}\BibitemShut
  {NoStop}%
\bibitem [{\citenamefont {Zheng}\ and\ \citenamefont {Chan}(2016)}]{Zheng2016}%
  \BibitemOpen
  \bibfield  {author} {\bibinfo {author} {\bibfnamefont {B.~X.}\ \bibnamefont
  {Zheng}}\ and\ \bibinfo {author} {\bibfnamefont {G.~K.~L.}\ \bibnamefont
  {Chan}},\ }\href@noop {} {\bibfield  {journal} {\bibinfo  {journal} {Phys.
  Rev. B}\ }\textbf {\bibinfo {volume} {93}},\ \bibinfo {pages} {035126}
  (\bibinfo {year} {2016})}\BibitemShut {NoStop}%
\bibitem [{\citenamefont {{\v{S}}imkovic~IV}\ \emph {et~al.}(2020)\citenamefont
  {{\v{S}}imkovic~IV}, \citenamefont {LeBlanc}, \citenamefont {Kim},
  \citenamefont {Deng}, \citenamefont {Prokof’ev}, \citenamefont
  {Svistunov},\ and\ \citenamefont {Kozik}}]{Simkovic2020}%
  \BibitemOpen
  \bibfield  {author} {\bibinfo {author} {\bibfnamefont {F.}~\bibnamefont
  {{\v{S}}imkovic~IV}}, \bibinfo {author} {\bibfnamefont {J.}~\bibnamefont
  {LeBlanc}}, \bibinfo {author} {\bibfnamefont {A.~J.}\ \bibnamefont {Kim}},
  \bibinfo {author} {\bibfnamefont {Y.}~\bibnamefont {Deng}}, \bibinfo {author}
  {\bibfnamefont {N.}~\bibnamefont {Prokof’ev}}, \bibinfo {author}
  {\bibfnamefont {B.}~\bibnamefont {Svistunov}},\ and\ \bibinfo {author}
  {\bibfnamefont {E.}~\bibnamefont {Kozik}},\ }\href@noop {} {\bibfield
  {journal} {\bibinfo  {journal} {Phys. Rev. Lett.}\ }\textbf {\bibinfo
  {volume} {124}},\ \bibinfo {pages} {017003} (\bibinfo {year}
  {2020})}\BibitemShut {NoStop}%
\bibitem [{\citenamefont {Ohno}(1964)}]{Ohno1964}%
  \BibitemOpen
  \bibfield  {author} {\bibinfo {author} {\bibfnamefont {K.}~\bibnamefont
  {Ohno}},\ }\href@noop {} {\bibfield  {journal} {\bibinfo  {journal} {Theor.
  Chim. Acta}\ }\textbf {\bibinfo {volume} {2}},\ \bibinfo {pages} {219}
  (\bibinfo {year} {1964})}\BibitemShut {NoStop}%
\bibitem [{\citenamefont {Chandross}\ and\ \citenamefont
  {Mazumdar}(1997)}]{Chandross1997}%
  \BibitemOpen
  \bibfield  {author} {\bibinfo {author} {\bibfnamefont {M.}~\bibnamefont
  {Chandross}}\ and\ \bibinfo {author} {\bibfnamefont {S.}~\bibnamefont
  {Mazumdar}},\ }\href@noop {} {\bibfield  {journal} {\bibinfo  {journal}
  {Phys. Rev. B - Condens. Matter Mater. Phys.}\ }\textbf {\bibinfo {volume}
  {55}},\ \bibinfo {pages} {1497} (\bibinfo {year} {1997})}\BibitemShut
  {NoStop}%
\bibitem [{\citenamefont {Bhattacharyya}\ \emph {et~al.}(2020)\citenamefont
  {Bhattacharyya}, \citenamefont {Rai},\ and\ \citenamefont
  {Shukla}}]{Bhattacharyya2020}%
  \BibitemOpen
  \bibfield  {author} {\bibinfo {author} {\bibfnamefont {P.}~\bibnamefont
  {Bhattacharyya}}, \bibinfo {author} {\bibfnamefont {D.~K.}\ \bibnamefont
  {Rai}},\ and\ \bibinfo {author} {\bibfnamefont {A.}~\bibnamefont {Shukla}},\
  }\href@noop {} {\bibfield  {journal} {\bibinfo  {journal} {J. Phys. Chem. C}\
  }\textbf {\bibinfo {volume} {124}},\ \bibinfo {pages} {14297} (\bibinfo
  {year} {2020})}\BibitemShut {NoStop}%
\bibitem [{\citenamefont {Bagley}\ and\ \citenamefont
  {Wornat}(2013)}]{Bagley2013}%
  \BibitemOpen
  \bibfield  {author} {\bibinfo {author} {\bibfnamefont {S.~P.}\ \bibnamefont
  {Bagley}}\ and\ \bibinfo {author} {\bibfnamefont {M.~J.}\ \bibnamefont
  {Wornat}},\ }\href {https://doi.org/10.1021/ef3021188} {\bibfield  {journal}
  {\bibinfo  {journal} {Energy and Fuels}\ }\textbf {\bibinfo {volume} {27}},\
  \bibinfo {pages} {1321} (\bibinfo {year} {2013})}\BibitemShut {NoStop}%
\bibitem [{\citenamefont {Eshuis}\ \emph {et~al.}(2010)\citenamefont {Eshuis},
  \citenamefont {Yarkony},\ and\ \citenamefont {Furche}}]{Eshuis2010}%
  \BibitemOpen
  \bibfield  {author} {\bibinfo {author} {\bibfnamefont {H.}~\bibnamefont
  {Eshuis}}, \bibinfo {author} {\bibfnamefont {J.}~\bibnamefont {Yarkony}},\
  and\ \bibinfo {author} {\bibfnamefont {F.}~\bibnamefont {Furche}},\
  }\href@noop {} {\bibfield  {journal} {\bibinfo  {journal} {J. Chem. Phys.}\
  }\textbf {\bibinfo {volume} {132}},\ \bibinfo {pages} {234114} (\bibinfo
  {year} {2010})}\BibitemShut {NoStop}%
\bibitem [{\citenamefont {Jeckelmann}\ and\ \citenamefont
  {White}(1998)}]{Jeckelmann1998}%
  \BibitemOpen
  \bibfield  {author} {\bibinfo {author} {\bibfnamefont {E.}~\bibnamefont
  {Jeckelmann}}\ and\ \bibinfo {author} {\bibfnamefont {S.~R.}\ \bibnamefont
  {White}},\ }\href@noop {} {\bibfield  {journal} {\bibinfo  {journal} {Phys.
  Rev. B}\ }\textbf {\bibinfo {volume} {57}},\ \bibinfo {pages} {6376}
  (\bibinfo {year} {1998})}\BibitemShut {NoStop}%
\bibitem [{\citenamefont {White}\ \emph {et~al.}(2020)\citenamefont {White},
  \citenamefont {Gao}, \citenamefont {Minnich},\ and\ \citenamefont
  {Chan}}]{White2020}%
  \BibitemOpen
  \bibfield  {author} {\bibinfo {author} {\bibfnamefont {A.~F.}\ \bibnamefont
  {White}}, \bibinfo {author} {\bibfnamefont {Y.}~\bibnamefont {Gao}}, \bibinfo
  {author} {\bibfnamefont {A.~J.}\ \bibnamefont {Minnich}},\ and\ \bibinfo
  {author} {\bibfnamefont {G.~K.~L.}\ \bibnamefont {Chan}},\ }\href@noop {}
  {\bibfield  {journal} {\bibinfo  {journal} {The J. Chem. Phys.}\ }\textbf
  {\bibinfo {volume} {153}} (\bibinfo {year} {2020})}\BibitemShut {NoStop}%
\bibitem [{\citenamefont {Mordovina}\ \emph {et~al.}(2020)\citenamefont
  {Mordovina}, \citenamefont {Bungey}, \citenamefont {Appel}, \citenamefont
  {Knowles}, \citenamefont {Rubio},\ and\ \citenamefont
  {Manby}}]{Mordovina2020}%
  \BibitemOpen
  \bibfield  {author} {\bibinfo {author} {\bibfnamefont {U.}~\bibnamefont
  {Mordovina}}, \bibinfo {author} {\bibfnamefont {C.}~\bibnamefont {Bungey}},
  \bibinfo {author} {\bibfnamefont {H.}~\bibnamefont {Appel}}, \bibinfo
  {author} {\bibfnamefont {P.~J.}\ \bibnamefont {Knowles}}, \bibinfo {author}
  {\bibfnamefont {A.}~\bibnamefont {Rubio}},\ and\ \bibinfo {author}
  {\bibfnamefont {F.~R.}\ \bibnamefont {Manby}},\ }\href@noop {} {\bibfield
  {journal} {\bibinfo  {journal} {Phys. Rev. Res.}\ }\textbf {\bibinfo {volume}
  {2}} (\bibinfo {year} {2020})}\BibitemShut {NoStop}%
\bibitem [{\citenamefont {Nusspickel}\ and\ \citenamefont
  {Booth}(2021)}]{nusspickel2021}%
  \BibitemOpen
  \bibfield  {author} {\bibinfo {author} {\bibfnamefont {M.}~\bibnamefont
  {Nusspickel}}\ and\ \bibinfo {author} {\bibfnamefont {G.~H.}\ \bibnamefont
  {Booth}},\ }\href@noop {} {\bibinfo {title} {Systematic improvability in
  quantum embedding for real materials}} (\bibinfo {year} {2021}),\ \Eprint
  {https://arxiv.org/abs/2107.04916} {arXiv:2107.04916 [cond-mat.str-el]}
  \BibitemShut {NoStop}%
\bibitem [{\citenamefont {Sandhoefer}\ and\ \citenamefont
  {Chan}(2016)}]{Sandhoefer2016}%
  \BibitemOpen
  \bibfield  {author} {\bibinfo {author} {\bibfnamefont {B.}~\bibnamefont
  {Sandhoefer}}\ and\ \bibinfo {author} {\bibfnamefont {G.~K.~L.}\ \bibnamefont
  {Chan}},\ }\href@noop {} {\bibfield  {journal} {\bibinfo  {journal} {Phys.
  Rev. B}\ }\textbf {\bibinfo {volume} {94}},\ \bibinfo {pages} {085115}
  (\bibinfo {year} {2016})}\BibitemShut {NoStop}%
\bibitem [{\citenamefont {Reinhard}\ \emph {et~al.}(2019)\citenamefont
  {Reinhard}, \citenamefont {Mordovina}, \citenamefont {Hubig}, \citenamefont
  {Kretchmer}, \citenamefont {Schollw{\"{o}}ck}, \citenamefont {Appel},
  \citenamefont {Sentef},\ and\ \citenamefont {Rubio}}]{Reinhard2019}%
  \BibitemOpen
  \bibfield  {author} {\bibinfo {author} {\bibfnamefont {T.~E.}\ \bibnamefont
  {Reinhard}}, \bibinfo {author} {\bibfnamefont {U.}~\bibnamefont {Mordovina}},
  \bibinfo {author} {\bibfnamefont {C.}~\bibnamefont {Hubig}}, \bibinfo
  {author} {\bibfnamefont {J.~S.}\ \bibnamefont {Kretchmer}}, \bibinfo {author}
  {\bibfnamefont {U.}~\bibnamefont {Schollw{\"{o}}ck}}, \bibinfo {author}
  {\bibfnamefont {H.}~\bibnamefont {Appel}}, \bibinfo {author} {\bibfnamefont
  {M.~A.}\ \bibnamefont {Sentef}},\ and\ \bibinfo {author} {\bibfnamefont
  {A.}~\bibnamefont {Rubio}},\ }\href@noop {} {\bibfield  {journal} {\bibinfo
  {journal} {J. Chem. Theory Comput.}\ }\textbf {\bibinfo {volume} {15}},\
  \bibinfo {pages} {2221} (\bibinfo {year} {2019})}\BibitemShut {NoStop}%
\end{thebibliography}
\providecommand{\noopsort}[1]{}\providecommand{\singleletter}[1]{#1}%

\appendix
\section{Proof for bounds on the number of cluster bosons} \label{App:Bounds}

    We prove that the maximum number of bosons coupled to the cluster is bounded by the number of particle-hole excitations in the cluster, which in general is at most $n_f^2$, where $n_f$ is the number of fragment degrees of freedom.
    \toadd{However, if the interaction kernel only couples a limited subset of particle-hole excitations in the RPA, then the following proof can be applied to that subset in isolation. This gives the upper bound on the boson number as the number of explicitly coupled particle-hole excitations allowed by the interaction kernel in the cluster, since all others are purely mean-field in character.
    More specifically, the initial RPA calculations in this work only couple between singlet particle-hole excitations, resulting in only $\frac{1}{4}n_f^2$ bosons. However, the inclusion of local exchange in the self-consistent EDMET interaction kernel couples all same-spin excitations, leading to $\frac{1}{2}n_f^2$ bosons as an upper bound. Only through inclusion also of spin-flip excitations (not considered in this work) would the number of bosons result in an $n_f^2$ scaling, as proven below.}
    

    This most general bound follows from the definition $\mat{S} = \mat{I} - (\mat{X}_{\mathrm{cl,f}}-\mat{Y}_{\mathrm{cl,f}})^T(\mat{X}_{\mathrm{cl,f}}+\mat{Y}_{\mathrm{cl,f}})$ and the fact that the \ac{RPA} closure relation must be satisfied within the fermionic portion of the cluster space, so $(\mat{X}_{\mathrm{cl,f}}-\mat{Y}_{\mathrm{cl,f}})(\mat{X}_{\mathrm{cl,f}}+\mat{Y}_{\mathrm{cl,f}})^T = \mat{I}_{{n_f}^2}$.
    Together, these lead to the relation
    \begin{align}
		(\mat{X}_{\mathrm{cl,f}} + \mat{Y}_{\mathrm{cl,f}}&)\mat{S} = \mat{0} \\
		&\left(= (\mat{X}_{\mathrm{cl,f}} + \mat{Y}_{\mathrm{cl,f}}) - \mat{I}_{{n_f}^2}(\mat{X}_{\mathrm{cl,f}} + \mat{Y}_{\mathrm{cl,f}})\right) \nonumber\\
		\mat{S}(\mat{X}_{\mathrm{cl,f}} - \mat{Y}_{\mathrm{cl,f}}&)^T = \mat{0} \\
		&\left(= (\mat{X}_{\mathrm{cl,f}} - \mat{Y}_{\mathrm{cl,f}})^T - (\mat{X}_{\mathrm{cl,f}} - \mat{Y}_{\mathrm{cl,f}})^T\mat{I}_{{n_f}^2}\right). \nonumber
	\end{align}
	These imply that all columns of $\mat{S}$ lie within the kernel of $(\mat{X}_{\mathrm{cl,f}} + \mat{Y}_{\mathrm{cl,f}})$, and all rows within the cokernel of $(\mat{X}_{\mathrm{cl,f}} - \mat{Y}_{\mathrm{cl,f}})^T$ (i.e. the kernel of $(\mat{X}_{\mathrm{cl,f}} - \mat{Y}_{\mathrm{cl,f}})$).
	Thus, the maximum number of linearly independent rows or columns of $\mat{S}$, and so its rank, is given by the minimum nullity between $(\mat{X}_{\mathrm{cl,f}} + \mat{Y}_{\mathrm{cl,f}})$ and $(\mat{X}_{\mathrm{cl,f}} - \mat{Y}_{\mathrm{cl,f}})$.
	Formally,
	
	\begin{equation}
	\textrm{rank}\left(\mat{S}\right) \leq \textrm{min}\left(\textrm{Null}\left(\mat{X}_{\mathrm{cl,f}} + \mat{Y}_{\mathrm{cl,f}}\right), \textrm{Null}\left(\mat{X}_{\mathrm{cl,f}} - \mat{Y}_{\mathrm{cl,f}}\right)\right).
	\label{eq:limit_rank_S}
	\end{equation}
	
	To complete the proof, we note that the closure relation requires that $(\mat{X}_{\mathrm{cl,f}} - \mat{Y}_{\mathrm{cl,f}})^T$ is the right inverse of $\mat{X}_{\mathrm{cl,f}} + \mat{Y}_{\mathrm{cl,f}}$.
	The existence of the right inverse implies both these matrices are of maximum possible rank (in this case $n_\textrm{f}^2$), since it requires all rows (columns) of $\mat{X}_{\mathrm{cl,f}} + \mat{Y}_{\mathrm{cl,f}}$ ($(\mat{X}_{\mathrm{cl,f}} - \mat{Y}_{\mathrm{cl,f}})^T$) be linearly independent.
	We can thus say
	\begin{equation}
		\textrm{rank}\left(\mat{X}_{\mathrm{cl,f}} + \mat{Y}_{\mathrm{cl,f}}\right) = \textrm{rank}\left(\mat{X}_{\mathrm{cl,f}} - \mat{Y}_{\mathrm{cl,f}}\right) = n_{f}^2,
		\label{eq:rank_XpY_XmY}
	\end{equation}
	so both relevant null spaces are of maximum size $n_f^2$, giving this as an upper bound on rank$(\mat{S})$. The definition of the bosons required to ensure the reproduction of these quantities ($\mat{X}_{\mathrm{cl,b}}$ and $\mat{Y}_{\mathrm{cl,b}}$) is given by 
	\begin{equation}
	    \mat{S}=(\mat{X}_{\mathrm{cl,b}}-\mat{Y}_{\mathrm{cl,b}})^T(\mat{X}_{\mathrm{cl,b}}+\mat{Y}_{\mathrm{cl,b}}),
	\end{equation}
	 which can be found from diagonalization of $\mat{S}$, bounding the number of coupled bosons in the cluster space by the rank$(\mat{S})$, given by $n_f^2$. However, we note that this is the formal bound on the number of bosons, and in practice the number of bosons in the results of this work are only half this number.

\section{EDMET Energy Estimator} \label{App:Energy}
    
    We derive a total energy functional which takes into account the mixed fermionic and bosonic nature of the bath space in the cluster calculation, and the origin of these contributions.
    We begin from the expression for the local energy arising from the total one- and two-body \acs{RDM} over the full system, given by
    \begin{equation}
        E_\text{frag} = \sum\limits_{p\in \text{frag}} \left(
        \sum\limits_q t_{pq} D_{qp}^{\text{tot}} +
        \frac{1}{2} \sum\limits_{qrs} \left(pq|rs\right) P_{qp|sr}^{\text{tot}}
        \right),
        \label{eq:lattice_energy_expression}
    \end{equation}
    where the one- and two-body \ac{RDMs} are defined as $D_{pq}^\text{tot} = \langle c_{q}^{\dagger} c_p \rangle$ and $P_{qp|sr}^\text{tot} = \langle c_p^{\dagger} c_r^{\dagger} c_s c_q \rangle$ respectively. We then rewrite the two-body energy contribution, separating the summations over $qrs$ into portions over the cluster (here defined as fragment and fermionic bath only) and environmental degrees of freedom, giving
    \begin{equation}
    \begin{aligned}
        &E_\text{frag}^\text{twobody} = \frac{1}{2}\sum\limits_{p\in \text{frag}} \left(\phantom{\sum\limits_{test}}\right.\\
        &\sum\limits_{q\in\text{cluster}} \left(
        \sum\limits_{rs\in \text{env}} \left(pq|rs\right) P_{qp|sr}^{\text{tot}}\right. + \sum\limits_{rs\in \text{cluster}} \left(pq|rs\right) P_{qp|sr}^{\text{tot}}\\
        &\left.+ \sum\limits_{r\in\text{cluster},s\in\text{env}} \left(\left(pq|rs\right) P_{qp|sr}^{\text{tot}} + \left(pq|sr\right) P_{qp|rs}^{\text{tot}}\right) \right)
        \\
        +&\sum\limits_{q\in\text{env}} \left(
        \sum\limits_{rs\in \text{env}} \left(pq|rs\right) P_{qp|sr}^{\text{tot}}\right. + \sum\limits_{rs\in \text{cluster}} \left(pq|rs\right) P_{qp|sr}^{\text{tot}}\\
        &\left.\left.+ \sum\limits_{r\in\text{cluster},s\in\text{env}} \left(\left(pq|rs\right) P_{qp|sr}^{\text{tot}} + \left(pq|sr\right) P_{qp|rs}^{\text{tot}}\right) \right)
        \right).
    \end{aligned}
    \label{eq:decomposed_lattice_Efrag}
    \end{equation}
    We canonicalize the environmental states, by ensuring that $D_{rs}^{\textrm{env}}$ is diagonal in the environment, denoting the resulting occupied states as $\{\alpha\}$, and the corresponding unoccupied environment states as $\{\beta\}$. Assuming a mean-field approximation to this environmental part of the wave function (the DMET approximation), this \ac{CAS} approximation modifies the two-body \ac{RDM} with one index $p$ constrained to be within the fragment, to become
    \begin{equation}
        P_{qp|sr}^\text{tot,CAS} = \left\{
        \begin{aligned}
            \phantom{-}D_{pq}^\text{frag}\delta_{sr} & \text{ if }q\in\text{cluster and } s \in\{j_\text{env}\} \\
            -D_{ps}^\text{frag}\delta_{qr} & \text{ if }s\in\text{cluster and } q \in\{j_\text{env}\} \\
            P_{qp|sr}^\text{frag} & \text{ if }pqrs\in\text{cluster}\\
            0 & \text{ otherwise.}
        \end{aligned}
        \right.
    \end{equation}

    The second of the eight terms in Eq.~\eqref{eq:decomposed_lattice_Efrag} corresponds to the correlated two-body local energy contribution of the cluster. 
    The first and last terms result in effective one-body mean-field contributions. 
    All other contributions are usually zero in this CAS approximation.

    The specific contributions to the effective one-body interaction are
    \begin{align}
        \frac{1}{2}\sum\limits_{\substack{p\in \text{frag}\\q\in\text{cluster}}}
        \sum\limits_{\substack{rs\in \text{env}\\ rs\in\{\alpha\}}}
        \left(pq|rs\right) P_{qp|sr}^{\text{tot}}
        &
        = \frac{1}{2}\sum\limits_{\substack{p\in \text{frag}\\q\in\text{cluster}}} D_{pq} \sum\limits_{i\in\{\alpha\}} \left(pq|ii\right)
        \label{eq:effective_onebody_coulomb}
        \\
        \frac{1}{2}\sum\limits_{\substack{p\in \text{frag}\\r\in\text{cluster}}}
        \sum\limits_{\substack{qs\in \text{env}\\ qs\in\{\alpha\}}}
        \left(pq|sr\right) P_{qp|rs}^{\text{tot}}
        &
        =
        -\frac{1}{2}\sum\limits_{\substack{p\in \text{frag}\\r\in\text{cluster}}} D_{pr} \sum\limits_{i\in\{\alpha\}} \left(pi|ir\right)
        \label{eq:effective_onebody_exchange}
    \end{align}
    for the first and last terms of Eq.~\eqref{eq:decomposed_lattice_Efrag}, leading to the overall \acs{DMET} energy expression
    \begin{equation}
    \begin{aligned}
        E_\text{f,frag} =& \sum\limits_{p\in \text{frag}} \left( \sum\limits_{q\in\text{clus}} \frac{t_{pq} + \tilde{h}_{pq}}{2} D_{pq}^\text{cl}\right. \\
        &\left. + \frac{1}{2} \sum\limits_{qrs\in\text{clus}} (pq|rs) P_{qp|sr}^\text{cl,f} \right)
    \end{aligned}
        \label{eq:Efrag_DMET}
    \end{equation}
    where the one- and two-body \acs{RDMs} within the cluster are $D_{pq}^\text{cl} = \langle c_{q}^{\dagger} c_p \rangle_{\text{cl}}$ and $P_{qp|sr}^\text{cl,f} = \langle c_p^{\dagger} c_r^{\dagger} c_s c_q \rangle_{\text{cl}}$ respectively, and defining the effective one-body interaction within a cluster as
    \begin{equation}
        \tilde{h}_{pq} = t_{pq} + \sum\limits_{rs} \left[(pq|rs) - (ps|rq) \right] D^\text{env}_{rs}.
        \label{eq:effective_onebody}
    \end{equation}
    
    We now improve this DMET energy expression, to go beyond the CAS (mean-field) environment description, and account for the bosonic RPA environmental fluctuations that are included via the bosonic bath contributions.
    We collect the portions of the first, third and fourth terms of Eq.~\eqref{eq:decomposed_lattice_Efrag} where $rs$ runs over ph-(de)excitations and at least one is purely environmental. This results in a two-body fragment energy contribution of the form
    \begin{equation}
    \begin{aligned}
        \frac{1}{2}
        \sum\limits_{\substack{p\in\text{frag}\\q\in\text{cluster}}}
        &\left(
        \sum\limits_{i\in\{\alpha\}}
        \sum\limits_{a}
        (\left(pq|ia\right) P_{qp|ai}^\text{tot} + \left(pq|ai\right) P_{qp|ia}^\text{tot})\right. \\
        &+
        \left.\sum\limits_{a\in\{\beta\}}
        \sum\limits_{i}
        (\left(pq|ia\right) P_{qp|ai}^\text{tot} + \left(pq|ai\right) P_{qp|ia}^\text{tot})
        \right)
        ,
    \end{aligned}
    \end{equation}
    which is assumed to be zero in DMET \toadd{due to the lack of inclusion of non-cluster correlated physics}.
    However, this environmental summation over the indices $i$ and $a$, where one of these indices must be in the environment, is exactly the fluctuation space from the cluster represented by the \ac{RPA}, represented by the bosonic bath space.
    As such, we can approximate the contribution over all environmental ph-(de)excitations within this expression as a summation over our bosonic excitations and de-excitations.
    This gives our final energy expression as
    \begin{align}
        E_\text{frag} \approx& E_\text{frag,f} + \frac{1}{2}\sum_{\substack{p\in \text{frag}\\ q\in\text{cluster}}} \sum_{n} ({\tilde V}_{pqn} P_{pq,n}^\text{cl,fb} + {\tilde V}_{pq\bar{n}}P_{pq,\bar{n}}^\text{cl,fb}) \label{eq:fbe1} \\
        =&E_\text{frag,f} + \frac{1}{2}\sum_{\substack{p\in \text{frag}\\ q\in\text{cluster}}} \sum_{n} ({\tilde V}_{pqn} P_{pq,n}^\text{cl,fb} + {\tilde V}_{qpn}P_{qp,n}^\text{cl,fb}),
        \label{eq:energy_expression}
    \end{align}
    where $P^\text{cl,fb}_{qp,n} = \langle c_p^{\dagger} c_q a_n \rangle_\textrm{cl}$ is the cluster fermion-boson \acs{RDM}, and $\bar{n}$ corresponds to a bosonic dexcitation index, where the simplification from Eq.~\ref{eq:fbe1} to Eq.~\ref{eq:energy_expression} arises due to hermiticity of the density matrices and hamiltonian. Note that this expression cannot easily be formulated as a simple modification to the two-body RDM, and also requires the projected coupling terms of the interaction to define.
    We should also note that this contribution is not double-counted, as it arises from correlated energy contributions in Eq.~\ref{eq:decomposed_lattice_Efrag} which are normally neglected. While the one-body effective contribution of the first term of Eq.~\ref{eq:decomposed_lattice_Efrag} is used to define Eqs~\eqref{eq:effective_onebody_coulomb}~\&~\eqref{eq:effective_onebody_exchange}, the bosonic contribution arises from the neglected two-body part. 

\end{document}